\documentclass[fleqn,usenatbib]{mnras}
\usepackage{newtxtext,newtxmath}
\usepackage[T1]{fontenc}
\usepackage{adjustbox}
\usepackage{color}
\definecolor{referee}{RGB}{0,0,0}
\DeclareRobustCommand{\VAN}[3]{#2}
\let\VANthebibliography\thebibliography
\def\thebibliography{\DeclareRobustCommand{\VAN}[3]{##3}\VANthebibliography}
\usepackage{ragged2e}
\newcommand{\ci}{[C\,{\sc i}]}
\usepackage{graphicx}	% Including figure files
\usepackage{amsmath}	% Advanced maths commands
\usepackage{lscape}
\title[Protocluster core around HerBS-70]{A dusty proto-cluster surrounding the binary galaxy HerBS-70 at $z=2.3$
}

\author[Tom Bakx et al.]{Tom J. L. C. Bakx$^{1,2,3}$\thanks{E-mail: tom.bakx@chalmers.se (Chalmers, Sweden)},
S. Berta$^{4}$,
H. Dannerbauer$^{5,6}$,
P. Cox$^{7}$,
K. M. Butler$^{4}$,
M. Hagimoto$^{2}$,\newauthor
D. H. Hughes$^{8}$,
D. A. Riechers$^{9}$,
P. P. van der Werf$^{10}$,
C. Yang$^{1}$,
A. J. Baker$^{11,12}$,
A. Beelen$^{13}$,\newauthor
G. J. Bendo$^{14}$,
E. Borsato$^{15}$,
V. Buat$^{16}$,
A. R. Cooray$^{17}$,
L. Dunne$^{18}$,
S. Dye$^{19}$,
S. Eales$^{18}$,\newauthor
R. Gavazzi$^{7,20}$,
A. I. Harris$^{21}$,
D. Ismail$^{16}$,
R. J. Ivison$^{22}$,
B. Jones$^{23}$,
M. Krips$^{4}$,
M. D. Lehnert$^{24}$,\newauthor
L. Marchetti$^{25,26}$,
H. Messias$^{27,28}$,
M. Negrello$^{18}$,
R. Neri$^{4}$,
A. Omont$^{7}$,
I. Perez-Fournon$^{29,5}$,\newauthor
A. Nanni$^{30,31}$,
N. Chartab$^{32,10,33}$,
S. Serjeant$^{34}$,
F. Stanley$^{4}$,
Y. Tamura$^{2}$,
S. A. Urquhart$^{34}$,\newauthor
C. Vlahakis$^{35}$,
A. Wei\ss$^{36}$,
and A. J. Young$^{11}$\\
Affiliations are listed at the end of the paper
}

\date{Accepted 2024 April 29. Received 2024 April 26; in original form 2022 December 23}

\pubyear{2022}
\usepackage{lipsum}
\begin{document}
\label{firstpage}
\pagerange{\pageref{firstpage}--\pageref{lastpage}}
\maketitle

\begin{abstract}
We report on deep SCUBA-2 observations at 850~\micron{} and NOEMA spectroscopic measurements at 2 mm of the environment surrounding the luminous, massive ($M_{*} \approx 2 \times 10^{11}$~M$_{\odot}$) {\it Herschel}-selected source HerBS-70.
This source was revealed by previous NOEMA observations to be a binary system of dusty star-forming galaxies at $z= 2.3$, with the East component (HerBS-70E) hosting an Active Galactic Nucleus (AGN). 
The SCUBA-2 observations detected, in addition to the binary system, twenty-one sources at $> 3.5 \sigma$ over an area of $\sim 25$~square comoving Mpc with a sensitivity of $1 \sigma_{850} = 0.75$~mJy.
The surface density of continuum sources around HerBS-70 is three~times higher than for field galaxies. The NOEMA spectroscopic measurements confirm the protocluster membership of three of the nine brightest sources through their CO(4--3) line emission, yielding a volume density 36 times higher than for field galaxies.
All five confirmed sub-mm galaxies in the HerBS-70 system have relatively short gas depletion times ($80 - 500$~Myr), indicating the onset of quenching for this protocluster core due to the depletion of gas.
% indicative of a possible protocluster core. However, across the wider environment the source counts do not appear to show any excess. 
% The strongest component of the HerBS-70 binary system, HerBS-70E, hosts an Active Galactic Nucleus traced by the radio emission, and five additional quasars are found in the overdense region surrounding HerBS-70.
% Additional nuclear activity is seen in the overdense region surrounding HerBS-70. %, and appears to be in excess from the sub-millimeter based star-formation rate for at least three nearby galaxies. 
The dark matter halo mass of the HerBS-70 system is estimated around $5 \times{} 10^{13}$~M$_{\odot}$, with a projected current-day mass of $10^{15}$~M$_{\odot}$, similar to the local Virgo and Coma clusters. 
These observations support the claim that DSFGs, in particular the ones with observed multiplicity, can trace cosmic overdensities.
% Spectroscopic confirmation of the redshifts of the galaxies surrounding the HerBS-70 binary system will provide essential information to further verify this candidate protocluster. 
\end{abstract}


\section{Introduction}
The evolution of the cosmic star-formation rate density (\citealt{madau2014}) paints a dynamic picture of galaxy evolution, with rapid stellar build-up at redshifts above 2, followed by a stark drop in star-formation rates of galaxies towards the present day. The origins of this down-turn are likely due to a mixture of feedback processes, broadly categorized as caused by {\it environmental} (strangulation, harassment and ram-pressure stripping)  and {\it internal} (star-formation and active-galactic nuclei) quenching effects \citep[e.g.,][]{Peng2015, Man2018,Walter2020}. 

Cosmic overdensities form pockets where the Universe is matter-dominated, and can trace galaxy evolution in more extreme environments, where the processes that characterize star-formation (i.e., fueling and feedback) happen on faster timescales and at earlier epochs \citep{Chiang2013,Muldrew2015,Contini2016,Overzier2016}.
Notably, models predict that the majority of high-redshift star formation occurs in overdense environments, where the compact regions drive gas away by environmental quenching through processes of gas stripping and the cessation of infalling gas that feeds star-formation. Meanwhile, the rapid evolution of galaxies can trigger Active Galactic Nuclei (AGN) which disrupt gas flow and star-formation from within \citep{Vayner2021}.

Hydrodynamical and parametric models of galaxy clusters predict that the intra-cluster medium (ICM) becomes too warm to readily collapse onto galaxies at redshifts 3 to 5, needing to form {\it cold streams} that fuel star-formation until $z \sim 2 - 3$ \citep[e.g.,][]{Dekel2006}. Cold flows could still persist further into the local Universe, however their extent might be more limited 
\citep{Dressler1980,Webb2015,Webb2017,Trudeau2019,Finner2020,H-L2020}. This generic picture \citep{Shimakawa2018} then predicts rapid quenching through a lack of fueling of the central galaxies, resulting in passive galaxy clusters at redshift 0 such as the $\sim$~10$^{15}$~M$_{\odot}$ Virgo cluster \citep{Fourque2001}. In particular, this sudden lack of inflowing gas in dense environments at $z \sim 2.5$ allows for a unique test of galaxy quenching \citep{Smail2024arXiv240108761S}.

Studies into the properties of these protocluster galaxies, however, find varying properties from protocluster to protocluster. They have measured metallicities deficient \citep{Valentino2015} or enhanced \citep{Shimakawa2015,Biffi2018} relative to field galaxies; they have star-formation rates boosted \citep{Shimakawa2018} or suppressed \citep{Tran2015} relative to field galaxies; they are more gas-rich than field galaxies \citep{Noble2017,Tadaki2019}, or show similar gas-richness \citep{Castignani2020,Lee2021}, or they show evidence of the re-distribution of metals across the galaxies through their circum-galactic medium \citep[CGM --][]{Wang2021}.

The galaxy populations within protoclusters appear diverse, and we are currently limited in our understanding through a sheer lack of identified high-$z$ protocluster environments. This diversity, however, could also be a selection bias, since protoclusters are found across the entire electromagnetic spectrum, from UV \citep{Steidel1998,Venemans2007,Toshikawa2012}, optical \citep{Hatch2011, Spitler2012} to radio \citep{Galametz2013}, each of which are likely sensitive to a specific galaxy population and therefore to a specific phase of cluster evolution.\footnote{An important exception to this are protoclusters selected by the Sunyaev-Zeldovich effect, which can accurately trace the total mass of dark matter haloes -- important for a cosmic census. However, this method becomes increasingly difficult to detect $z > 1$ protoclusters (e.g., \citealt{DiMascolo2021,Meinke2021}).} 

To extend these studies out to higher redshifts, various rare populations of galaxies or quasars have been proposed to trace overdense regions, with mixed results. For example, quasars have been shown to trace protoclusters by \cite{Kim2009,Utsumi2010,Morselli2014,Kikuta2017,Kikuta2019,Ota2018,GarciaVergara2022}, however the studies by \citealt{Banados2013,Mazzucchelli2017,Goto2017} failed to reveal any overdense environments, suggesting that this method is not necessarily robust \citep[e.g.,][]{Champagne2018}.

Dusty star-forming galaxies (DSFGs) \citep{blain2002,casey2014,hodge2020} are known to trace some of the most active star-forming regions in the Universe, with very short gas depletion timescales \citep{swinbank2014,aravena2016coldgasmass,Canameras2018}. Observational studies have shown a strong connection between these galaxies and their wider {\color{referee} CGM} \citep{Banerji2011,Dannerbauer2014, Emonts2016,Emonts2018,Spilker2020,Spilker2020b,berta2021,Butler2021,Riechers2021} rapidly enriching their environments while also showcasing examples of gas feeding at $z \sim 2 - 3$. 
Far-infrared cameras have successfully identified such protocluster systems, requiring multi-hour observations to reveal a handful to tens of components \citep{Ivison2000,Stevens2003,Stevens2005,Stevens2010,deBreuck2004,Chapman2009,Tamura2009,Aravena2010,Capak2011,Casey2013Scuba2,Dannerbauer2014,Miller2018,Zeballos2018,ConnorSmith2019,GG2019,Lacaille2019,Guaita2022,Zhang2022}. 
%could more efficiently target protocluster environments, and in turn, we could provide early indications to whether quenching should be sought inside or outside of galaxies.
The small number of sources revealed by continuum cameras limit the statistical power of sub-mm observations (e.g., \citealt{Lewis2018}), however it is important to characterize these sources, as they dominate the star-formation in these environments \citep[e.g.,][]{Casey2013Scuba2,Dannerbauer2014,Bussmann2015,Miller2018,Oteo2018,Long2020,Stach2021}. Stacking experiments by \cite{Kubo2019} and \cite{Alberts2021} further show an excess of mid- and far-infrared emission from protoclusters in {\it WISE} and {\it Planck} data. 
Environmental studies of DSFGs also help characterize the evolution of the massive end of star-formation, as DSFGs are the likely progenitors of massive ellipticals, which could help identify potential environmental effects in order to explain recent observations of massive quenched galaxies being found at $z \sim 4$ \citep{Straatman2014,Glazebrook2017,Simpson2017,Schreiber2018,Looser2023}. 
Meanwhile, simulations suggest that source multiplicity might be a better indicator than the more time-variable measure of star-formation rate \citep{Remus2022}. In order to efficiently identify overdensities, tracing out the environments around distant DSFGs with neighbouring sources, we would address both strong selection criteria for overdense environments. %, and it could make for an effective selection criteria towards the largest protoclusters. 

%The results here presented report on deep SCUBA-2 observations at 850 microns of the field surrounding the Herschel galaxy H-ATLAS J130140.2+292918 (HerBS-70), which was revealed to be a binary system of DSFGs at z=2.3 \citep{neri2019}. The two galaxies, HerBS-70E and W, are separated by 140~kpc, with HerBS-70E hosting an Active Galactic Nuclei (AGN) has recently found by Stanley et al., subm. 

Follow-up observations of luminous high-redshift galaxies detected in large surveys such as Herschel, in particular using sub-millimetre facilities, have enabled to identify candidate proto-clusters by resolving the sources in multiple galaxies at similar redshift. A recent example is the z-GAL survey that measured, using the NOrthern Extended Millimetre Array (NOEMA), precise spectroscopic redshifts of 135 dusty, bright \textit{Herschel}-selected galaxies \citep{Cox2023,neri2019}. In some cases, the fields displayed multiplicity and revealed candidate proto-clusters. One example is the Herschel galaxy H-ATLAS J130140.2+292918 (HerBS-70), that was revealed by the NOEMA observations to be a binary system of dusty star-forming galaxies at $z=2.3$, where the two components are separated by $16.5”$ and the East component (HerBS-70E) hosts an Active Galactic Nucleus (AGN - \citealt{neri2019,Stanley2023}). 

In this work, we report on deep continuum observations at 850 \micron{} of the environment around HerBS-70 using SCUBA-2 on the James Clerk Maxwell Telescope (JCMT) and targeted NOEMA spectroscopic measurements of the most likely protocluster candidates in the field. 
% In this work, we report on deep SCUBA-2 observations at 850 microns on the James Clerk Maxwell Telescope (JCMT) and subsequent NOEMA spectroscopic follow-up of the most likely protocluster candidates in the field surrounding the {\it Herschel} galaxy H-ATLAS J130140.2+292918 (HerBS-70). %Using multi-wavelength data, we can then identify nearby associated galaxies and look for evidence of an overdense region surrounding HerBS-70. This source was revealed to be a binary system of DSFGs at $z=2.3$ \citep{neri2019}. %The two galaxies, HerBS-70E and W, are separated by 140~kpc, with HerBS-70E hosting an Active Galactic Nuclei (AGN) has recently found by Stanley et al., subm. 
% These properties make HerBS-70 a potential candidate to be at the center of an overdense region and was therefore chosen as a pilot source for exploring with SCUBA-2 its larger scale environment in sub-mm colours. 
In Section~\ref{sec:Target}, we describe the properties of HerBS-70, and why it is a good candidate lighthouse for a cosmic overdensity.
We then report on the continuum map observations using the SCUBA-2 instrument on the JCMT and spectroscopic observations using NOEMA at 2~mm, including the reduction steps, flux extraction, completion and flux boosting estimates in Section~\ref{sec:results}. 
Section~\ref{sec:discussion} characterizes the overdensity in HerBS-70 using the spectroscopic and continuum observations. We continue our discussion on the {\it Herschel} photometry to produce photometric redshifts, and evaluates the continuum overdensity estimates based on source counts and spatial source distribution. 
Subsequently, this Section places the wider environment around HerBS-70 in its cosmological context, and looks forward to expanded studies of the environments surrounding dusty star-forming galaxies in general. 
We provide our conclusions in Section~\ref{sec:conclusions}.
Throughout this paper, we assume a flat $\Lambda$-CDM cosmology with the best-fit parameters derived from the \textit{Planck} results \citep{Planck2020}, which are $\Omega_\mathrm{m} = 0.315$, $\Omega_\mathrm{\Lambda} = 0.685$ and $h = 0.674$.

\section{HerBS-70: A rapidly-quenching, AGN-hosting source with a nearby companion}
\label{sec:Target}

% \subsection{Identification and characterization of the central source, HerBS-70}

\begin{table*}
    \caption{Sub-mm derived properties of HerBS-70E \& W}
    \centering
        \begin{tabular}{lrp{0.1in}lrp{0.1in}ll}\hline
                                        & \multicolumn{3}{c}{HerBS-70E}  & \multicolumn{3}{c}{HerBS-70W} \\ \hline
L$_{IR}$ (10$^{11}$ $L_{\odot}$)        & $180.0$&$\pm$&$6.0$        & $42.8$&$\pm$&$5.0 $        & \citet{Berta2023}\\ 
SFR ($M_{\odot}$/yr)                    & $2000$&$\pm$&$100 $     & $470$&$\pm$&$80$       & \citet{Berta2023}\\
                                        % & $3800$&$\pm$&$200 $     & $880$&$\pm$&$200$       & \citet{Stanley2023} \\
M$_{\rm dust}$ (10$^{10}$ $M_{\odot}$)  & $0.42$&$\pm$&$0.04$        & $0.10$&$\pm$&$0.04$        & \citet{neri2019}\\ 
M$_{\rm gas}$ (10$^{11}$ $M_{\odot}$)   & $3.3$&$\pm$&$0.9$         & $3.1$&$\pm$&$0.5$         & \citet{Berta2023}\\ 
                                        & $ 2.7$&$\pm$&$0.6$           & $1.5$&$\pm$&$0.3 $         & \citet{Stanley2023}\\ 
t$_{\rm depl}$ (10$^9$ yr)              & $0.17$&$\pm$&$0.06$        & $0.66$&$\pm$&$0.11$        & \citet{Berta2023} \\
                                        % & $0.7$&$\pm$&$0.2$          & $1.7$&$\pm$&$0.9$          & \citet{Stanley2023} \\
                                        \hline
    \label{tab:herbs70}
\end{tabular}
\end{table*}

% NOEMA observations by \cite{neri2019} targeted 13 bright {\it Herschel} targets, and two of these (HerBS-70 and HerBS-95) showed signs of multiplicity . We used this multiplicity as a potential tracer for overdense regions, and we chose HerBS-70 (H-ATLAS J130140.2+292918) as a pilot source for examining the larger-scale environment in sub-mm colours. 
As part of a redshift campaign of bright, dusty galaxies identified by {\it Herschel}-ATLAS \citep{eales2010,bakx18,Bakx2020Erratum}, NOEMA observations of HerBS-70 revealed a non-lensed binary system at $z_{\rm spec} = 2.3$ \citep{neri2019,Cox2023}. Table~\ref{tab:herbs70} summarizes the sub-mm properties of the HerBS-70 system. % with wide (FWHM~$\sim$770 km/s; $z_{\rm spec} = 2.3077$) CO emission lines, with narrow (FWHM~$\sim$140 km/s; $z_{\rm spec} = 2.3155$) CO lines reported by \cite{neri2019}. 
The two galaxies of the HerBS-70 system, HerBS-70W and HerBS-70E, are separated by a projected distance of $16.5”$ (140 kpc at $z = 2.3$), and the spectroscopic redshift difference between the galaxies is relatively small, corresponding to a velocity shift of $350$~km/s ($z_{\rm spec} = 2.3077$ and 2.3155 for HerBS-70E and -70W, respectively; \citealt{neri2019}). The source HerBS-70E displays strong double-peaked emission lines of CO (3--2) and (4--3) with widths of $\sim 770$~km/s. In contrast, HerBS-70W has significantly narrower ($\sim 140$~km/s) single-peaked emission lines. Subsequent observations using the VLA measured the CO(1--0) emission line of the HerBS-70 binary system and revealed a bright radio continuum at 30 GHz in HerBS-70E indicating the presence of a radio luminous AGN in this source \citep{Stanley2023}.
% The galaxies are separated by a projected 16.5~arcseconds (140~kpc at $z=2.3$), and the spectroscopic redshift difference between the galaxies ($z_{\rm spec} =2.3077$ and 2.3155 for HerBS-70E and -70W, respectively) corresponds to 5~comoving Mpc along the radial direction if the velocity shift (350 km/s) is solely due to cosmological expansion.
% Subsequent observations using the VLA measured the CO(1--0) and the associated radio continuum of the HerBS-70 binary system (\citealt{Stanley2023}; ). 
Based on these results, the following properties were derived for HerBS-70E and W. The 8-1000 microns infrared luminosities are $1.8 \times 10^{13}$ and $4 \times 10^{12}$ L$_{\odot}$ for HerBS-70 E and W, respectively, yielding star formation rates of 2000 and 470~M$_{\odot}$/yr, assuming a \cite{Chabrier2003} initial mass function \citep{Berta2023}, and dust masses of 4 and $1 \times 10^9$ M$_{\odot}$ \citep{Draine2007dustmodel,neri2019}.  The molecular gas masses of HerBS-70E and W are $2.7 \times 10^{11}$ and $1.5 \times 10^{11}$ M$_{\odot}$, with estimated depletion time scales of 170 and 660 Myr, respectively, although these values vary strongly between different gas-mass and star-formation estimates \citep{Berta2023,Stanley2023}. %In addition, the VLA observations...... in this source. 
HerBS-70E poses an interesting case of feedback through multiple mechanisms. Firstly, the detection of a buried AGN in HerBS-70E suggests ongoing quenching through negative feedback \citep{Stanley2023}. Secondly, star-formation feedback is likely present, given the excessive $\sim 1500-3000$~M$_{\odot}$/yr star-formation rate of HerBS-70E \citep{RowanRobinson2016}. Finally, the depletion timescale (i.e., the star-forming gas mass divided by the star-formation rate) is on the order of 170~Myr \citep{swinbank2014,Bakx2020IRAM,Stanley2023,Berta2023}, which places a strong limit on the quenching timescale in the case of no inflowing gas. 

\begin{figure}
    \centering
    \includegraphics[height=\linewidth,angle=270]{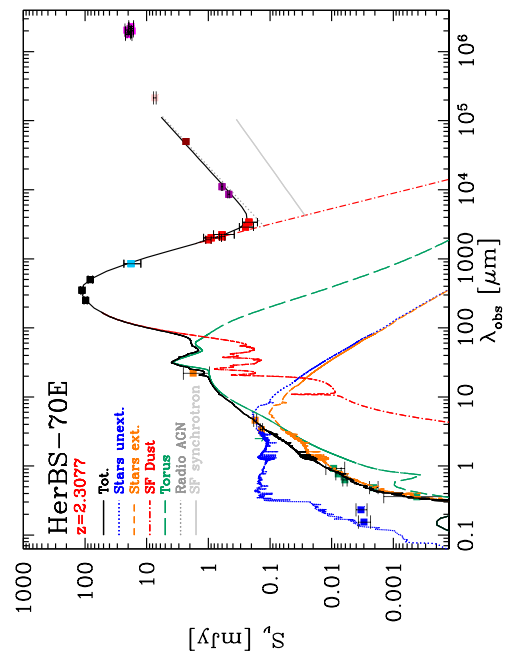}
    \caption{The UV-to-radio SED fit of HerBS-70E shows different emission components across the various wavelength ranges. The sub-mm emission is dominated by the dust obscuring star-formation, while the AGN component is dominant at near-infrared and radio wavelengths. The fit is performed using the SED3FIT code \citep{Berta2013}, and uses data from GALEX (UV; {\it blue squares}), SDSS (optical; {\it green squares}), UKIDSS (near-infrared; {\it green squares}), WISE (mid-infrared; {\it orange squares}), {\it Herschel} H-ATLAS (far-infrared; {\it black squares}), SCUBA-2 (this paper {\it light-blue squares}), NOEMA (mm; \citealt{neri2019}; {\it dark-red squares}), VLA (radio; {\it purple squares}), FIRST and LOFAR (m-wavelength -- \citealt{Stanley2023}; {\it brown, salmon and pink squares}). The expected radio emission based on the FIR-radio correlation of \citet{delhaize2017} at the redshift of HerBS-70E is shown as a {\it grey line} beyond 10$^4$~\micron{}, and is well below the observed radio flux densities. The excess emission is due to AGN contribution ({\it dotted grey line}). Similarly at the mid-infrared, excess emission in WISE is fit by a power-law-like SED indicative of a dusty torus at the centre of HerBS-70.
    }
    \label{fig:spectrum}
\end{figure}

In order to better understand HerBS-70E and its potential for being a central source in an overdense region, we reproduced the UV-to-radio spectral energy distribution (SED) with a composite model that includes the emission of stars, of dust heated by stellar light and by an AGN dusty torus (see Fig.~\ref{fig:spectrum}).
We adopted the SED3FIT code \citep{Berta2013}, which combines \cite{Bruzual2003} stellar models, dust emission powered by star formation \citep[see also][]{cunha2008}, and an AGN torus component \citep{Feltre2016, Feltre2012, Fritz2006}, fitted simultaneously to the observed SED.

The radio continuum is not directly fitted by the code, but it is instead computed from the model IR luminosity adopting the radio-FIR correlation at the redshift of HerBS-70E \citep[][see also \citealt{Magnelli2015}]{delhaize2017}. The expected radio emission due to star formation (grey solid line of Figure~\ref{fig:spectrum}) is much weaker than the observed radio continuum fluxes. This excess confirms the presence of an AGN in the central regions of HerBS-70E (see also \citealt{Stanley2023}). The WISE data reveal a mid-IR excess in the broad-band SED, that is fitted with an AGN torus (see green long-dashed line in Fig.~\ref{fig:spectrum}).

An additional radio power law (dotted grey line in Fig.~\ref{fig:spectrum}) is added a posteriori to reproduce the radio emission of the HerBS-70E central AGN. At the lowest frequencies, the LOFAR data suggest a change of slope of the radio emission. This change of slope is mainly due to synchrotron self absorption (SSA), but note that the low-frequency spectral index is shallower than that of pure SSA, thus indicating that also other physical processes influence it.

The total $L(\rm IR, 8-1000\mu\textrm{m})$ of the source is between 1.6 and 2.9$\times10^{13}$ L$_\odot$, with the contribution of star formation amounting between 1.4 and $2.6\times10^{13}$ L$_\odot$. The latter corresponds to a SFR in the range $\sim 1500-2800$  M$_\odot$/yr, adopting the \citet{kennicutt1998} conversion after correcting it to the \citet{Chabrier2003} IMF.

The inclusion of the AGN component is important to estimate the stellar mass, because the near-infrared emission of HerBS-70E turns out to be dominated by the dusty torus ($\sim 60$\% between 1.0 and 2.5 $\mu$m and up to $\sim90$\% between 2.5 and 5.0 $\mu$m).
The resulting best-fit stellar mass is 2.0$^{+0.4}_{-0.5} \times 10^{11}$~M$_{\odot}$. This suggests a low specific star-formation rate (SFR per unit stellar mass or a stellar-mass doubling timescale) of $70-130$~Myr$^{-1}$.   %, depending on the assumed initial-mass function.
Combined with the available molecular gas from CO emission, this implies a gas fraction on the order of $\sim 60$ per cent \citep[see also][]{Ismail2023,Stanley2023,neri2019}, in line with soon-to-be quenched galaxies \citep{Gobat2018,Gobat2020} at both low and high redshifts.
 
Although no conclusive evidence yet exists, there are several indications that point to the presence of a cosmic overdensity \citep{Overzier2016}:
{\em i.} HerBS-70E is forming stars at a high rate ($\sim 1500-3800$~M$_{\odot}$/yr), while having an additional galaxy in its vicinity that could indicate an overdensity of objects within its environment (HerBS-70W). 
{\em ii.} Meanwhile, HerBS-70E has already built up a significant stellar mass at less than three billion years after the Big Bang, which would mean an average star-formation rate of at least $\sim 70$~M$_{\odot}$/yr. These star-formation rates in the early Universe are primarily seen in overdense regions in the early Universe \citep{Chiang2017}. 
{\em iii.} The stellar assembly further required a large amount of gas and/or efficient feeding of galaxy to reach $2 \times{} 10^{11}$~M$_{\odot}$, and HerBS-70E is currently rapidly processing all remaining gas within its system ($t_{\rm dep} = 170$~Myr). Although the presence of an AGN can drive quenching -- and is predicted to amply exist in $z = 2$ protoclusters \citep{Shimakawa2018} -- it is also possible that galaxies in its environment are removing gas flows needed for future star formation. Since many of these indications are found together in the HerBS-70 system, a characterization of its environment can teach us both about the evolution of massive galaxies and about large-scale structures in the Universe.

% \section{Observations and Data Reduction}
\section{Observations and results}
\label{sec:results}
In order to find evidence that the binary system HerBS-70 is part of a protocluster, we first carried out deep 850 \micron{} continuum observations using the SCUBA-2 sub-millimetre camera to search for dusty star-forming galaxies in the Mpc-scale environment around HerBS-70. These observations were followed-up by targeted spectroscopic measurements with NOEMA of selected sources identified in the 850 \micron{} continuum map in order to derive their spectroscopic redshifts and verify that the values are around $z \sim 2.3$, comparable to HerBS-70. In this Section, we discuss the SCUBA-2 and NOEMA observations and results, including data reduction, flux extraction steps, as well as produce estimates of the completeness and flux boosting.

\subsection{Continuum map with SCUBA-2}
\label{sec:continuum}
The mapping speed of sub-mm spectroscopic instruments is relatively low, which favours a two-step process in identifying overdense regions. First, a fast continuum mapping observation characterizes dusty galaxies, and these sources can then be observed through targeted spectroscopic follow-up. Here, we use SCUBA-2 on JCMT to look for dusty galaxies in the vicinity of the HerBS-70 system.

\subsubsection{SCUBA-2 850 micron observations}
\begin{figure*}
\includegraphics[width=\linewidth]{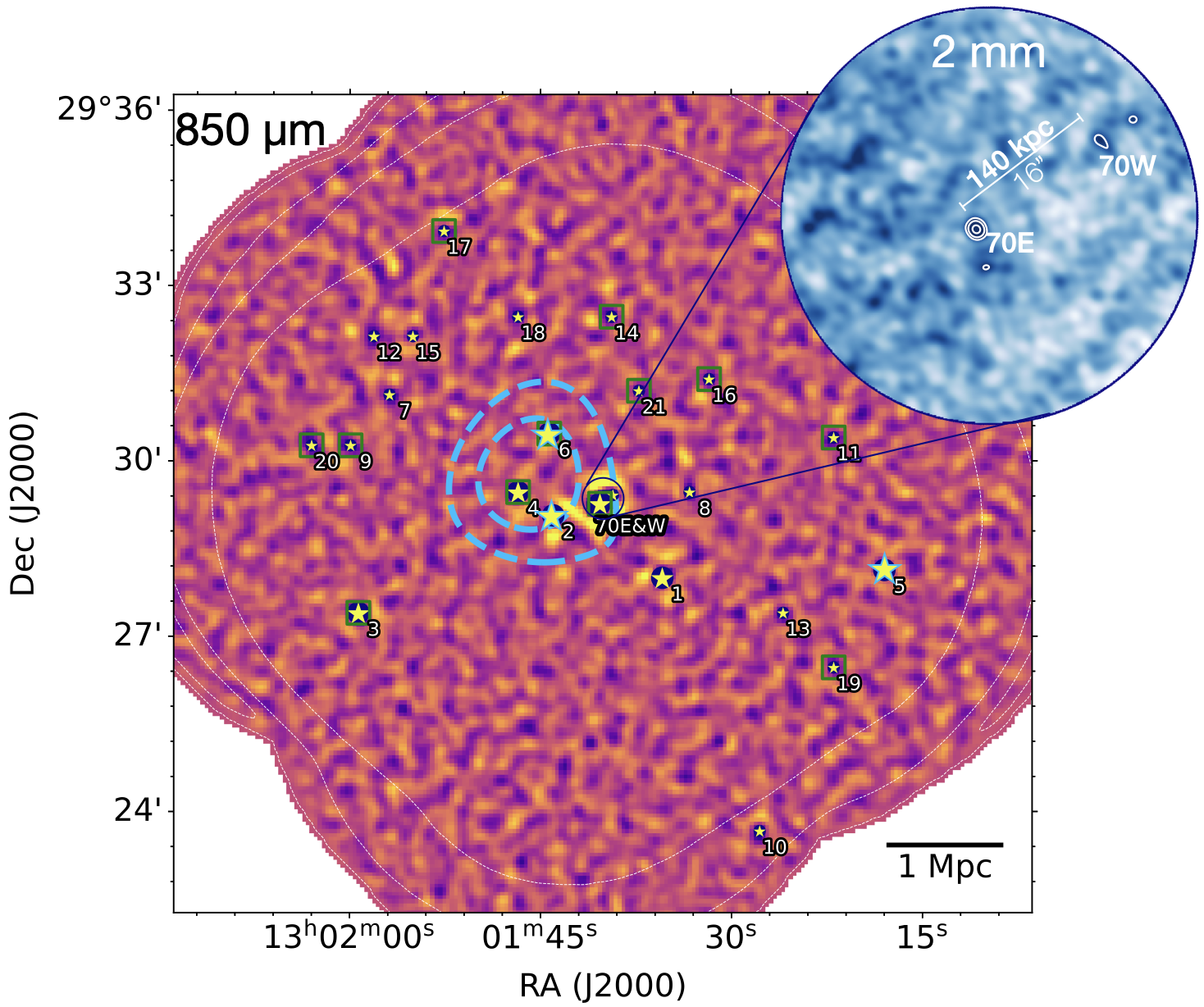}
    \caption{The SCUBA-2 850~\micron{} signal-to-noise ratio map of a $14 \times{} 14$ arcmin$^2$ area centered on HerBS-70, equivalent to $7 \times{} 7$ Mpc at $z=2.3$. In total, seven sources are detected at $>5 \sigma$ ({\it large stars}), including the binary system HerBS-70 (unresolved by SCUBA-2), and fifteen sources are detected at $>3.5 \sigma$ ({\it small stars}). {\color{referee} The spectroscopically-confirmed sources from the NOEMA observations are identified by {\it stars with blue borders}, i.e., H70.2, .5 and .6. The sources H70.1, .3, .4, .7, .8 and .10 were found to not be close in redshift to HerBS-70.} {\it Thin dashed contours} indicate the flux standard deviation at 2, 5 and 10~mJy. The {\it blue dashed contours} indicate the regional overdensity on the scales of $\sim 1$ arcminute ($\sim 0.5$~Mpc) of the map, based on field number count statistics from \citet{geach2017} which account for the inhomogeneous survey depth (see Section~\ref{sec:protoclustercore}). Sources with nearby {\it Herschel} identifications are shown with {\it green boxes} (Section~\ref{sec:herschelcounterparts}). Four sources at $> 5 \sigma$ are located within 1 Mpc of the central source at a surface density in excess of 2.9 times that of field galaxies (see text for details). This overdensity suggests HerBS-70 is located at the centre of a protocluster core.
    The inset figure shows the 150~GHz continuum emission ({\it background and contours}) centered on the position of HerBS-70E, with HerBS-70W at 140~kpc (16~arcseconds) projected distance. Both sources have robust spectroscopic redshifts from detections of CO(1--0), CO(3--2) and CO(4--3) emission lines at $z=2.307$ and 2.315, respectively (\citealt{neri2019} and \citealt{Stanley2023}). No additional sources are located behind the inset figure. }
    \label{fig:mapOfHerBS70}
\end{figure*}
HerBS-70 was observed using the SCUBA-2 bolometer instrument mounted on the JCMT. The instrument is able to observe both 450 and 850~$\mu$m emission simultaneously through a dichroic mirror \citep{Holland2013}. 
Through a {\sc DAISY} pattern scanning observations, HerBS-70 was observed following a continuous petal-like track, providing a central 3 arcminute region of uniform exposure time, and keeping one part of the array on-source at all times \citep{Chapin2013}.
The observations were conducted under project code M20BP049 (PI: T. Bakx) in excellent weather conditions ($\tau_{\rm 225} \approx 0.04 - 0.06$) on the 29th of January, the 1st and 3rd of February 2021 for a total of 4.9 hours of on-source time.

\subsubsection{Data reduction}
We followed the \textit{zero masking} data reduction prescription described in \cite{Holland2017}, which performs the standard deep observation pipeline using the {\sc starlink}'s {\sc SMURF} and {\sc KAPPA} packages. This observation pipeline is supplemented with the addition of masking out the central bright source that was previously detected with {\it Herschel} and NOEMA.
The data is reduced iteratively using the {\sc mapmaker} algorithm, which removes noise in order to produce a final astronomical estimate in the image plane. This assumed astronomical signal is then removed from the time-series data, and the noise estimations are done again until a threshold noise level is reached. 

The resulting astronomical map is then \textit{matched filtered} to the point spread function (PSF) of the JCMT, assumed to be 13~arcseconds. The assumed flux-conversion-factor is taken to be 537~Jy/pW/beam \citep{Smith2019,Mairs2021}. The signal-to-noise ratio (SNR) map is produced using the {\sc kappa} recipe {\sc makesnr}. At the centre of the map, the noise level is $0.75$~mJy/beam, which approaches the background-confusion limit due to the far-infrared background caused by unresolved sub-mm emitters within the beam of the JCMT. 

Fluxes are extracted using the matched-filtered signal-to-noise map using the peak emission value. Sources are differentiated by emission peaks separated by more than one beam (13 arcseconds), and each source should have at least $3.5 \sigma$. None of the sources are extended beyond the JCMT beam. Noise estimates are made directly from the noise map, which is also produced by the {\sc makemap} algorithm. 

Figure~\ref{fig:mapOfHerBS70} shows the resulting map of around 14 by 14 arcminutes, with all sources above $3.5 \sigma$ marked as stars, with sources $>5 \sigma$ displayed with larger stars. Thin dashed lines indicate the regions where the background noise is below 2, 5 and 10~mJy. In total, 21 SCUBA-2 sources are identified at $> 3.5 \sigma$ in addition to HerBS-70 (the E and W components are not resolved in the SCUBA-2 image), and are labeled according to decreasing SNR ratio (Table~\ref{tab:sources}). 
An inset of the NOEMA 2-mm continuum map is shown in the top-right corner (from \citealt{neri2019}), where HerBS-70E and 70W are seen as resolved sources. Green squares show the sources with nearby {\it Herschel} galaxies identified through the HELP catalogue\footnote{http://herschel.sussex.ac.uk} \citep{Shirley2021HELP}. The {\it light-blue} contours show the regions where the field is denser than thrice the expected number of background sources (see Section~\ref{sec:protoclustercore}).

\begin{figure}
    \centering
    \includegraphics[width=\columnwidth]{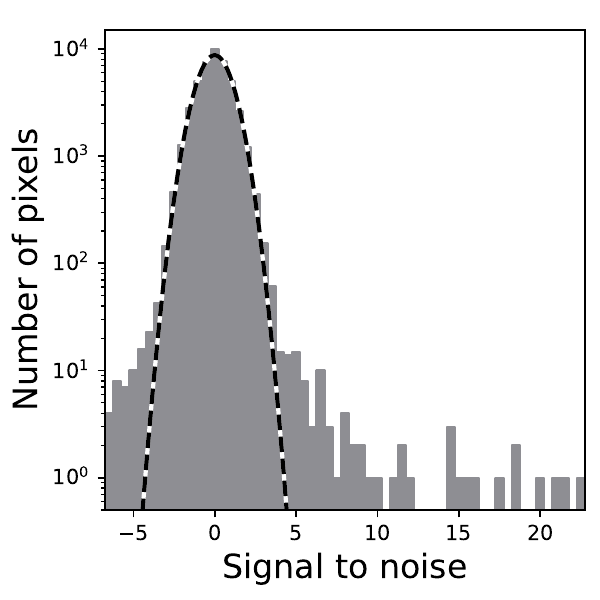}
    \caption{The signal-to-noise ratio distribution of the map in Figure~\ref{fig:mapOfHerBS70}. The majority of the data points follow the expected Gaussian ($\sigma = 1$) distribution, however the excess of bins at the positive end reflect the detection of multiple sources. At the negative tail, the excess of bins are caused by the bright nature of HerBS-70, where the negative component of the PSF of the JCMT causes deep troughs in the SNR map \citep{Dempsey2013,geach2017}}
    \label{fig:SNR}
\end{figure}
Figure~\ref{fig:SNR} shows the distribution of SNR values of each pixel in the map. Broadly, the bulk of the pixels follow the behaviour expected from white noise, i.e., a Gaussian fit with $\sigma = 1$ fits the bulk of the data well. Several peaks at the low- and high-end of the SNR distribution indicate the existence of sources. The negative pixels are caused by the PSF of the JCMT beam, which has negative rings around the peak of $\sim20$ per cent of the peak flux. Excess noise pixels around $\sim -5 \sigma$ are thus expected given the peak SNR of around 20$\sigma$.

% \subsubsection{Flux extraction}

% \subsection{Results and Analysis}
% \label{sec:results}
% In this section, we correct for known issues in extracting low signal-to-noise data from continuum maps, namely the source catalog completeness, contamination and flux boosting. We further compare the extracted sources to the sub-mm maps from {\it Herschel}.

\subsubsection{Completeness and Flux Boosting}
In order to maximise the scientific gain of the SCUBA-2 image, we assessed the contamination (i.e., false-positives) and completeness (i.e., false-negatives) of our SCUBA-2 observations, and corrected for the Eddington bias in the flux boosting. The Eddington bias accounts for the observed upward boost of fluxes, since there are more faint galaxies that scatter upwards than there are rarer bright galaxies that scatter towards fainter fluxes. Here, we broadly follow the procedure documented in \cite{Eales2000,Cowie2002, Scott2002, Borys2003, Coppin2006, Lewis2018}, and in line with recent deep SCUBA-2 observations such as the SCUBA-2 Cosmological Legacy Survey \citep{geach2017} and the SCUBA-2 survey of COSMOS \citep{Casey2013Scuba2, Simpson2019}.

We use the negative map of the SCUBA-2 image as a reference image without source emission. We first process this map by removing any spurious identifications above 2.5$\sigma$ and below -2.5$\sigma$, and proceed to add fake sources based on the source count distribution from \cite{weiss2009}, assuming the typical FWHM of the beam (13 arcseconds). We execute this process 5000 times and extract our galaxies through their peak point-source flux, which we then compare it against the original {\it injected} flux. We then use the noise maps to translate the {\it injected} fluxes into SNR estimates. 

\begin{figure}
	\includegraphics[width=\columnwidth]{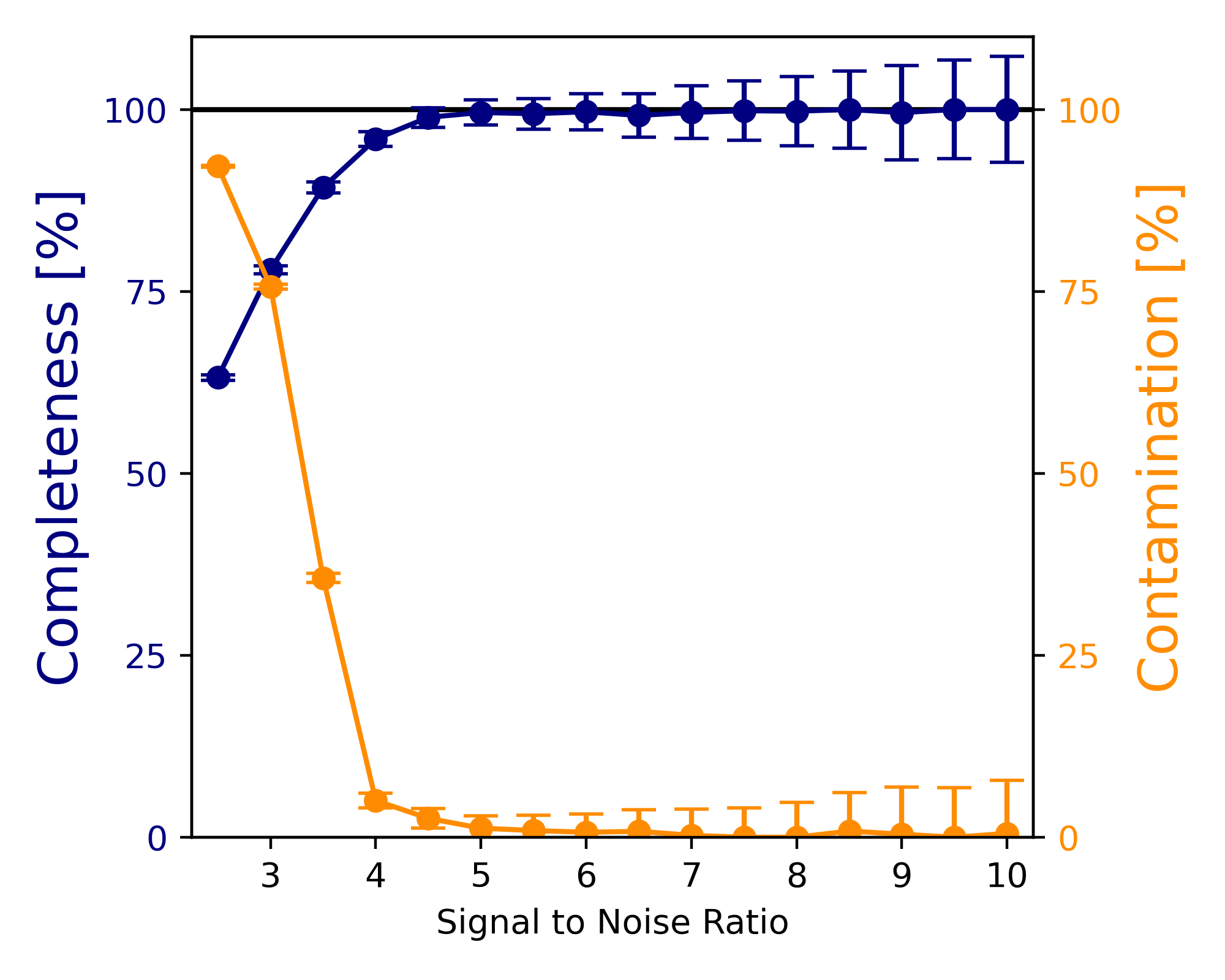}
    \caption{{\it Blue markers} indicate the false-negative rate of the SCUBA-2 extracted sources within each SNR bin. Sources extracted around 3$\sigma$ are around 75 per cent complete. {\it Orange markers} show the false-positive rate of the SCUBA-2 extracted sources within each SNR bin. Around 75 per cent of the sources at 3$\sigma$ are expected to be contaminants, however the comparison of these low-fidelity sources with {\it Herschel} counterparts can improve the purity of this sample. }
    \label{fig:completion}
\end{figure}

Figure~\ref{fig:completion} compares the {\it injected} sources within each SNR bin to the fraction of sources that were actually extracted, as shown in the blue points {\it completeness} graph, i.e., the false-negatives. Even down to 3$\sigma$, we are able to extract 75 per cent of the sources. The {\it orange contamination} graph indicates the fraction of spurious sources that are extracted, i.e, the false-positives. The errors on both estimates are proportional to the square root of the number of simulated sources in each signal-to-noise ratio bin. The number of contaminants increases rapidly below $4\sigma$, even suggesting that up to 75 per cent of $3\sigma$ sources are spurious. As a balance between contamination and completion, we take a $3.5 \sigma$ cut as an extraction threshold.

\begin{figure}
	\includegraphics[width=\columnwidth]{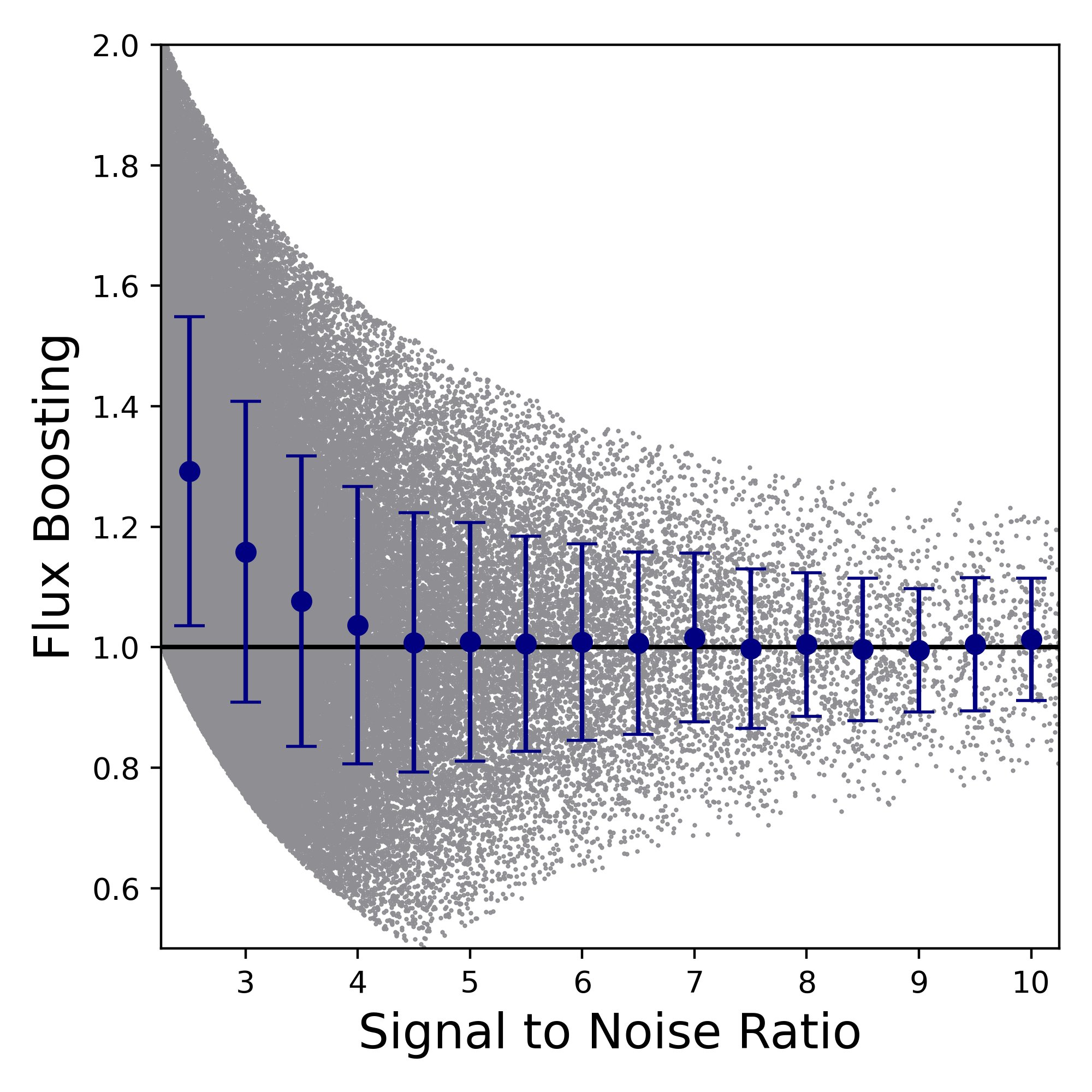}
    \caption{The flux boosting (i.e., the extracted / injected flux ratio) at 850~\micron{} is estimated from the fake injected sources ({\it grey dots}), where the detected faint sources likely have fainter fluxes. The {\it blue errorbars} indicate the averaged flux boosting within SNR bins with a width of 0.5 $\sigma$. We adjust for the flux boosting for the 850~\micron{} fluxes in Table~\ref{tab:sources}, and include the additional uncertainty to the instrumental uncertainty in quadrature.}
    \label{fig:fluxboosting}
\end{figure}

Figure~\ref{fig:fluxboosting} shows the Eddington boosting depending on the SNR of a source. The Eddington boosting is calculated as the ratio between the measured and the injected flux. Galaxies detected at lower signal-to-noise are more likely to be boosted in flux due to the exponential nature of the source count distribution, with the faintest sources typically appearing $\sim20$ per cent brighter than their true flux. We therefore appropriately adjust for the flux boosting for the 850~\micron{} fluxes in Table~\ref{tab:sources}, and include the additional uncertainty to the instrumental uncertainty in quadrature.

% \include{table_with_sourcefluxes}
% \resizebox{\textwidth}{!}{
\begin{table*}
    \caption{Observed and derived properties of sources detected at 850\micron{} in the field surrounding HerBS-70}
    \label{tab:sources}
    % \centering
    \begin{tabular}{rccrcccccccccc} \hline
\# & RA & DEC & SNR & $z_{\rm phot}$  & SFR$_{850}$ & $L_{\rm IR}$ & $M_{\rm dust}$ & $S_{\rm 250 \mu m}$& $S_{\rm 350 \mu m}$& $S_{\rm 500 \mu m}$& $S_{\rm 850 \mu m}$' & Bias & $\mathcal{F}$ \\ 
& [hms] & [dms] & & & [M$_{\odot}$/yr]& [$10^{12} L_{\odot}$]& [$10^9 M_{\odot}$] & [mJy]& [mJy]& [mJy]& [mJy] \\
\hline
70EW  & 13:01:40.3 & +29:29:16 & 22.6 & 1.8 $\pm$ 0.4 & 1800 & 10.2 & 5.3 & 122.3 $\pm$ 6.1 & 139.3 $\pm$ 5.6 & 106.4 $\pm$ 6.0 & 16.9 $\pm$ 1.9 & 1.00 & 0.0\\
70.S1 & 13:01:35.4 & +29:28:00 & 8.9 & 3.4 $\pm$ 0.6 & 780 & 4.6 & 2.4 & 11.9 $\pm$ 6.1$^{\dagger}$ & 12.6 $\pm$ 5.6$^{\dagger}$ & 15.7 $\pm$ 6.0$^{\dagger}$ & 7.5 $\pm$ 1.2 & 1.00 & 0.0\\
70.S2 & 13:01:44.0 & +29:29:04 & 6.8 & 2.0 $\pm$ 0.4 & 550 & 3.2 & 1.7 & 23.3 $\pm$ 6.1$^{\dagger}$ & 32.4 $\pm$ 5.6$^{\dagger}$ & 28.7 $\pm$ 6.0$^{\dagger}$ & 5.3 $\pm$ 1.1 & 1.01 & 0.0\\
70.S3 & 13:02:00.3 & +29:27:24 & 6.6 & 1.7 $\pm$ 0.4 & 750 & 4.4 & 2.3 & 54.0 $\pm$ 6.0 & 57.5 $\pm$ 5.5 & 40.3 $\pm$ 6.3 & 7.2 $\pm$ 1.5 & 1.01 & 0.0\\
70.S4 & 13:01:46.7 & +29:29:28 & 6.5 & 1.7 $\pm$ 0.4 & 580 & 3.4 & 1.8 & 34.2 $\pm$ 6.1 & 37.4 $\pm$ 5.6 & 20.6 $\pm$ 6.1 & 5.6 $\pm$ 1.2 & 1.01 & 0.0\\
70.S5 & 13:01:18.0 & +29:28:08 & 5.1 & 2.6 $\pm$ 0.5 & 670 & 3.9 & 2.0 & 16.3 $\pm$ 6.1$^{\dagger}$ & 19.5 $\pm$ 5.6$^{\dagger}$ & 12.8 $\pm$ 6.0$^{\dagger}$ & 6.5 $\pm$ 1.8 & 1.01 & 0.0\\
70.S6 & 13:01:44.3 & +29:30:28 & 5.0 & 2.0 $\pm$ 0.4 & 430 & 2.5 & 1.3 & 17.9 $\pm$ 6.0 & 19.8 $\pm$ 5.3 & 6.5 $\pm$ 6.3 & 4.2 $\pm$ 1.2 & 1.01 & 0.0\\
70.S7 & 13:01:56.9 & +29:31:08 & 4.7 & 3.5 $\pm$ 0.6 & 540 & 3.1 & 1.6 & 6.2 $\pm$ 6.1$^{\dagger}$ & 9.5 $\pm$ 5.6$^{\dagger}$ & 12.5 $\pm$ 6.0$^{\dagger}$ & 5.2 $\pm$ 1.6 & 1.01 & 0.0\\
70.S8 & 13:01:33.3 & +29:29:28 & 4.3 & 2.8 $\pm$ 0.5 & 400 & 2.3 & 1.2 & 6.1 $\pm$ 6.1$^{\dagger}$ & 13.2 $\pm$ 5.6$^{\dagger}$ & 6.1 $\pm$ 6.0$^{\dagger}$ & 3.9 $\pm$ 1.2 & 1.02 & 0.0\\
70.S9 & 13:02:00.0 & +29:30:16 & 4.3 & 1.3 $\pm$ 0.3 & 520 & 3.0 & 1.6 & 58.4 $\pm$ 5.9 & 46.2 $\pm$ 5.9 & 25.8 $\pm$ 6.5 & 5.0 $\pm$ 1.6 & 1.02 & 0.0\\
70.S10 & 13:01:27.8 & +29:23:40 & 4.2 & 3.6 $\pm$ 0.6 & 1100 & 6.6 & 3.4 & 13.1 $\pm$ 6.1$^{\dagger}$ & 19.0 $\pm$ 5.6$^{\dagger}$ & 16.7 $\pm$ 6.0$^{\dagger}$ & 10.9 $\pm$ 3.5 & 1.02 & 0.0\\
70.S11 & 13:01:22.0 & +29:30:24 & 4.1 & 0.7 $\pm$ 0.2 & 460 & 2.7 & 1.4 & 102.5 $\pm$ 6.0 & 57.1 $\pm$ 5.2 & 29.5 $\pm$ 6.2 & 4.4 $\pm$ 1.5 & 1.03 & 0.0\\
70.S12 & 13:01:58.1 & +29:32:08 & 3.9 & 3.4 $\pm$ 0.6 & 480 & 2.8 & 1.5 & 5.9 $\pm$ 6.1$^{\dagger}$ & 8.2 $\pm$ 5.6$^{\dagger}$ & 12.3 $\pm$ 6.0$^{\dagger}$ & 4.7 $\pm$ 1.6 & 1.05 & 0.1\\
70.S13 & 13:01:25.9 & +29:27:24 & 3.8 & 2.6 $\pm$ 0.5 & 430 & 2.5 & 1.3 & 8.4 $\pm$ 6.1$^{\dagger}$ & 17.6 $\pm$ 5.6$^{\dagger}$ & 29.3 $\pm$ 6.0$^{\dagger}$ & 4.2 $\pm$ 1.5 & 1.05 & 0.2\\
70.S14 & 13:01:39.4 & +29:32:28 & 3.7 & 0.7 $\pm$ 0.2 & 510 & 3.0 & 1.6 & 128.5 $\pm$ 6.2 & 76.3 $\pm$ 5.4 & 35.9 $\pm$ 5.7 & 5.0 $\pm$ 1.8 & 1.06 & 0.2\\
70.S15 & 13:01:55.0 & +29:32:08 & 3.7 & 2.4 $\pm$ 0.4 & 420 & 2.5 & 1.3 & 12.0 $\pm$ 6.1$^{\dagger}$ & 16.7 $\pm$ 5.6$^{\dagger}$ & 22.0 $\pm$ 6.0$^{\dagger}$ & 4.1 $\pm$ 1.5 & 1.06 & 0.2\\
70.S16 & 13:01:31.7 & +29:31:24 & 3.7 & 2.3 $\pm$ 0.4 & 370 & 2.2 & 1.1 & 16.5 $\pm$ 5.8 & 5.7 $\pm$ 5.8 & 3.6 $\pm$ 6.4 & 3.6 $\pm$ 1.3 & 1.06 & 0.2\\
70.S17 & 13:01:52.6 & +29:33:56 & 3.7 & 2.1 $\pm$ 0.4 & 630 & 3.7 & 1.9 & 21.2 $\pm$ 5.8 & 18.2 $\pm$ 5.7 & 8.8 $\pm$ 6.0 & 6.1 $\pm$ 2.2 & 1.06 & 0.3\\
70.S18 & 13:01:46.8 & +29:32:28 & 3.7 & 1.7 $\pm$ 0.4 & 400 & 2.3 & 1.2 & 23.4 $\pm$ 6.1$^{\dagger}$ & 17.2 $\pm$ 5.6$^{\dagger}$ & 10.1 $\pm$ 6.0$^{\dagger}$ & 3.9 $\pm$ 1.4 & 1.06 & 0.3\\
70.S19 & 13:01:22.0 & +29:26:28 & 3.6 & 1.5 $\pm$ 0.3 & 480 & 2.8 & 1.4 & 30.3 $\pm$ 6.0 & 20.4 $\pm$ 5.5 & 9.9 $\pm$ 6.0 & 4.6 $\pm$ 1.7 & 1.07 & 0.3\\
70.S20 & 13:02:03.0 & +29:30:16 & 3.5 & 1.6 $\pm$ 0.3 & 410 & 2.4 & 1.2 & 38.0 $\pm$ 5.8 & 31.3 $\pm$ 5.6 & 27.3 $\pm$ 5.7 & 4.0 $\pm$ 1.5 & 1.08 & 0.4\\
70.S21 & 13:01:37.3 & +29:31:12 & 3.5 & 1.9 $\pm$ 0.4 & 310 & 1.8 & 0.9 & 13.6 $\pm$ 5.6 & 16.5 $\pm$ 5.7 & 3.7 $\pm$ 5.9 & 3.0 $\pm$ 1.1 & 1.08 & 0.4\\
    \hline \end{tabular}
\raggedright \justify \vspace{-0.2cm}
\textbf{Notes:} 
Col. 1: The source identification of the sources shown in Figure~\ref{fig:mapOfHerBS70}. 
Col. 2 \& 3: The right ascension and declination of the sources.
Col. 4: The signal-to-noise ratio of the sources from the map, extracted from the peak pixel using the raw signal and noise of the sources, i.e., uncorrected for Eddington bias that is included in column 12.
Col. 5: Photometric redshift derived using the two-temperature modified black-body from \cite{pearson13}. 
Col. 6: The bolometric infrared luminosity ($8-1000$\micron{}), derived from the deboosted 850$\mu$m flux density assuming $z = 2.3$. We assume a single-temperature modified black-body at 30~K.
Col. 7: The star-formation rate derived from the FIR luminosity-to-SFR conversion factor, $1.73 \times 10^{-10}\ \mathrm{M_{\odot} yr^{-1}/L_{\odot}}$, is valid for a Salpeter $1-100\ \mathrm{M_{\odot}}$ IMF \citep{kennicutt1998}. Note that these SFR estimates are based on a single-temperature estimate, and thus the HerBS-70E\&W SFR estimate deviates from the full fit provided in \citet{bakx18,Bakx2020Erratum}.
Col. 8: The dust mass estimate assuming a single-temperature modified black-body at 30~K at $z = 2.3$, assuming $\kappa_{d} = 10.41$~cm$^2$/g at 1900~GHz, and scaling according to $\kappa_{d}(\nu/\nu_{\rm ref})^{\beta}$ with $\beta = 2$ following \cite{draine2003}.
Col. 9, 10 \& 11: The {\it Herschel} photometry derived from the HELP catalogue. When unavailable (indicated by a \,$^{\dagger}$), we extract the fluxes directly from the {\it Herschel} maps (\url{www.h-atlas.org}) and use the errors documented in \cite{valiante2016}. 
Col. 12: The 850~\micron{} fluxes and errors, corrected for flux boosting, with the errors including both the instrumental and observational biases.
Col. 13: The flux boosting bias.
Col. 14: The false-positive probability, as a probability between 0 and 1.
\end{table*}
% }

\subsection{Spectroscopic characterization with NOEMA}
The continuum sub-mm camera observations have revealed a tentative overdensity in the central region surrounding HerBS-70 (see Section~\ref{sec:continuumoverdensityestimates}). Subsequent spectroscopic confirmation requires targeted deep observations to reveal potential spectroscopic lines. Here, we use NOEMA at 2mm to target the CO(4--3) and \ci{} emission lines simultaneously to identify companion sources at a similar redshift to the HerBS-70 system, and for those companions, measure their molecular gas masses.
\label{sec:spectroscopicFollowUp}

\subsubsection{NOEMA observations of CO(4--3) and atomic carbon}
\begin{figure}
    \centering
    \includegraphics[width=0.8\linewidth]{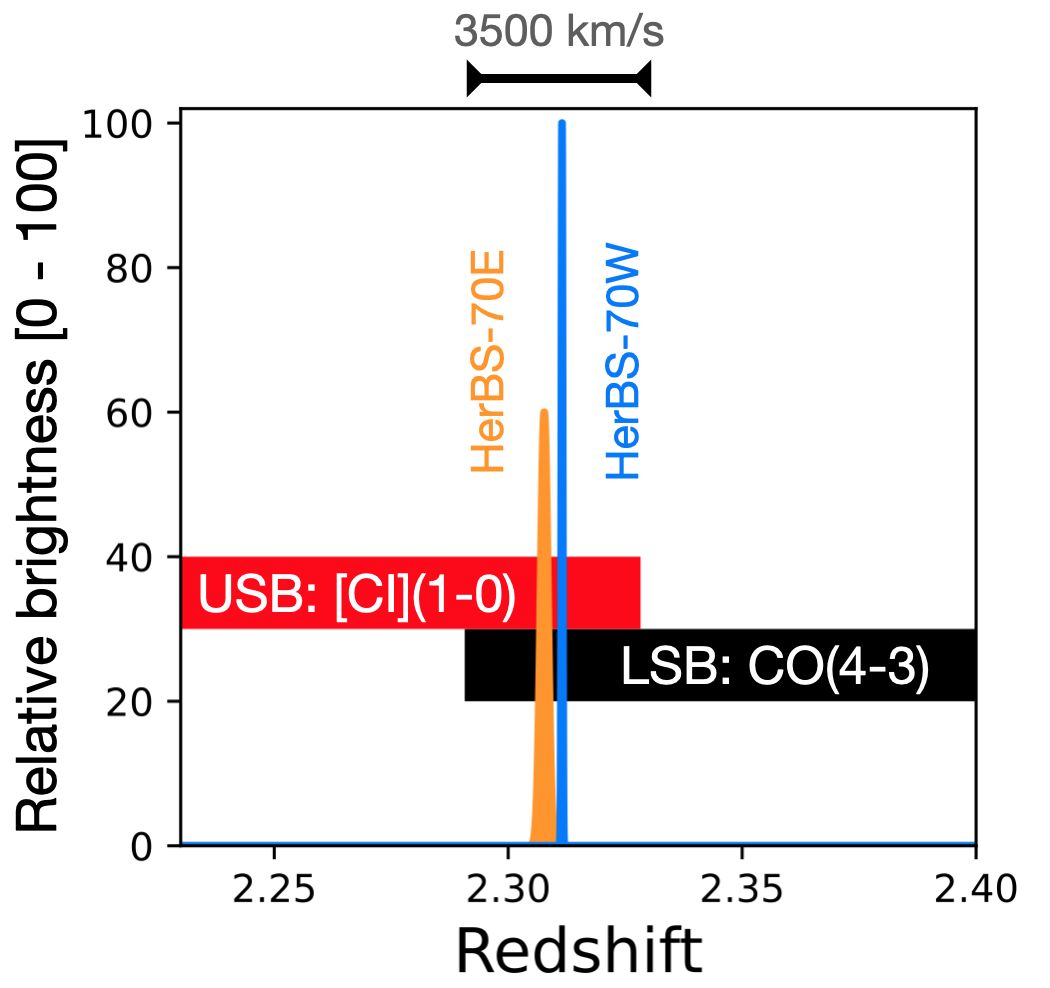}
    \caption{The relative emission of CO(4--3) of HerBS-70E (\textit{orange}) and HerBS-70W (\textit{blue}) are shown against the selected PolyFix set-ups, where the CO(4--3) and \ci{} lines can be targeted in the LSB and USB, respectively. In total, the observations probe roughly $3500$~km/s around HerBS-70E.}
    \label{fig:NOEMA_setup}
\end{figure}

During the summer of 2023, the NOEMA interferometer observed nine of the brightest protocluster member candidates in track-sharing mode (S23CI001; PI: Bakx). 
The source H70.9 was excluded, as it lies close to a known quasar and because of its low photometric redshift ($z_{\rm phot} = 1.3 \pm 0.3$; Table~\ref{tab:sources}).
The PolyFix receivers are set up to cover 132.925 to 140.625~GHz (LSB) and 148.404 to 156.104~GHz (USB), with the expected frequency for CO(4--3) is at 139.43~GHz and for \ci{} is at 148.71~GHz. Figure~\ref{fig:NOEMA_setup} depicts the chosen set-up to include both CO(4--3) and \ci{} emission at the expected frequency of HerBS-70, and includes roughly $\pm 2000$~km/s around this system, wide enough to include most of the protocluster members \citep{Kurk2000,Koyama2013,Koyama2021}. The 2~mm band is sensitive to a brighter part of the continuum emission of protocluster members surrounding HerBS-70 and facilitate the identification of the expected position of the line emission, instead of an alternative strategy that would target CO(3--2) in the 3~mm band where the dust continuum of nearby galaxies is likely too faint to be detected.

\subsubsection{Data reduction and flux extraction}
Phase and amplitude calibrators were chosen in close proximity to the track-shared sources in order to reduce uncertainties in astrometric and relative amplitude calibration. MWC349 and 3C273 were used as absolute flux calibrators. The absolute flux calibration was estimated to be accurate to within 10\%. The data were calibrated, averaged in polarisation, imaged, and analysed in the GILDAS software package\footnote{\url{http://www.iram.fr/IRAMFR/GILDAS/}}.
The $uv$-tables were produced from calibrated visibilities in the standard way and cleaned using natural weighting and support masks defined on each of the detected sources. Continuum maps were produced for each sideband, excluding the spectral ranges that include emission lines. Continuum extraction was performed on cleaned continuum maps, after correcting for primary beam attenuation. The resulting beam size is 3.6 by 3.4~arcseconds, and the continuum depth is between 60 and 90~$\mu$Jy/beam per side-band, while the $1 \sigma$ spectral sensitivity is between 1.3 to 1.8~mJy/beam in a 35~km/s bin. The spectra of each source were extracted from cleaned cubes within polygonal apertures defined by hand on the emission line channels to enhance the signal-to-noise of the spectral detection. 

\begin{table*}
    \caption{NOEMA fluxes of protocluster candidates}
    \label{tab:NOEMAfluxes}
    % \centering
    \begin{tabular}{rccccccccc} 
    \hline
    Source & RA & Dec & $S_{\rm LSB}$ & $S_{\rm USB}$ & $z_{\rm spec}$ & $\Delta V$ & $\delta V_{\rm H70}$ & $S dV$ CO(4--3) & $S dV$ \ci{} \\ 
     & [hms] & [dms] & [mJy] & [mJy] &  & [km/s] & [km/s] & [Jy km / s] & [Jy km / s] \\
    \hline
    
    H70.S1 & 13:01:35.29 & 29:28:03.2 & 0.43 $\pm$ 0.08 & 0.71 $\pm$ 0.12 & \\
    H70.S2 & 13:01:44:19 & 29:29:03.4 & 0.29 $\pm$ 0.07 & 0.40 $\pm$ 0.09 & 2.3100 & 980 $\pm$ 160 & -205 & 2.35 $\pm$ 0.39 & $< 1.53$ \\
    H70.S3 & & & $< 0.20$ & $< 0.27$ \\
    H70.S4 & & & $< 0.22$ & $< 0.29$ \\ 
    H70.S5 & 13:01:18.08 & 29:28:14.0 & $0.39 \pm 0.09$ & $0.62 \pm 0.11$ & 2.2895 &  $86 \pm 22$ & $- 1665$ & $0.47 \pm 0.12$ & $< 0.46$ \\ 
    H70.S6 & 13:01:44:43 & 29:30:32.0 & $< 0.24$ & $0.36 \pm 0.09$ & 2.3294 & $160 \pm 36$  & 1947  & $0.62 \pm 0.14$  & - \\
    H70.S7 & & & $< 0.26$ & $< 0.27$  &\\ 
    H70.S8 & & & $< 0.20$ & $< 0.25$ \\ 
    H70.S10 & & & $< 0.21$ & $< 0.29$\\
        \hline
        \end{tabular}
\raggedright \justify \vspace{-0.2cm}
\textbf{Notes:} 
Col. 1: Source name.
Col. 2 \& 3: The right ascension and declination of the sources based on NOEMA data.
Col. 4 \& 5: The Lower and Upper Side-Band continuum fluxes or $3 \sigma$ upper limits.
Col. 6: The spectroscopic redshift associated with the line identification.
Col. 7: The full-width at half-maximum of the line. 
Col. 8: The velocity-separation between the source and HerBS-70.
Col. 9 \& 10: The line fluxes and $3 \sigma$ upper limits.
\end{table*}

\begin{figure*}
    \centering
    \includegraphics[width=0.8\linewidth]{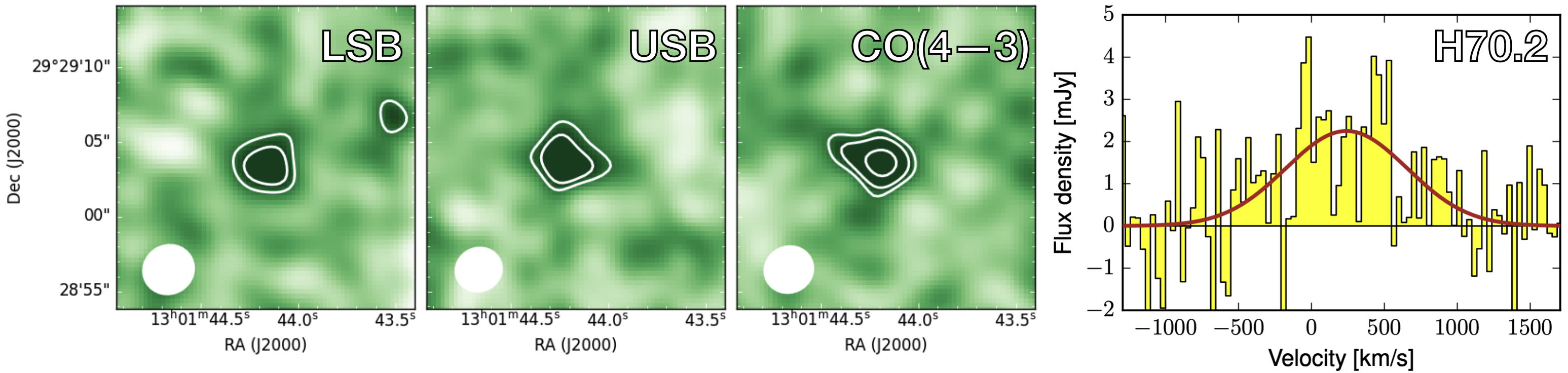}
    \includegraphics[width=0.8\linewidth]{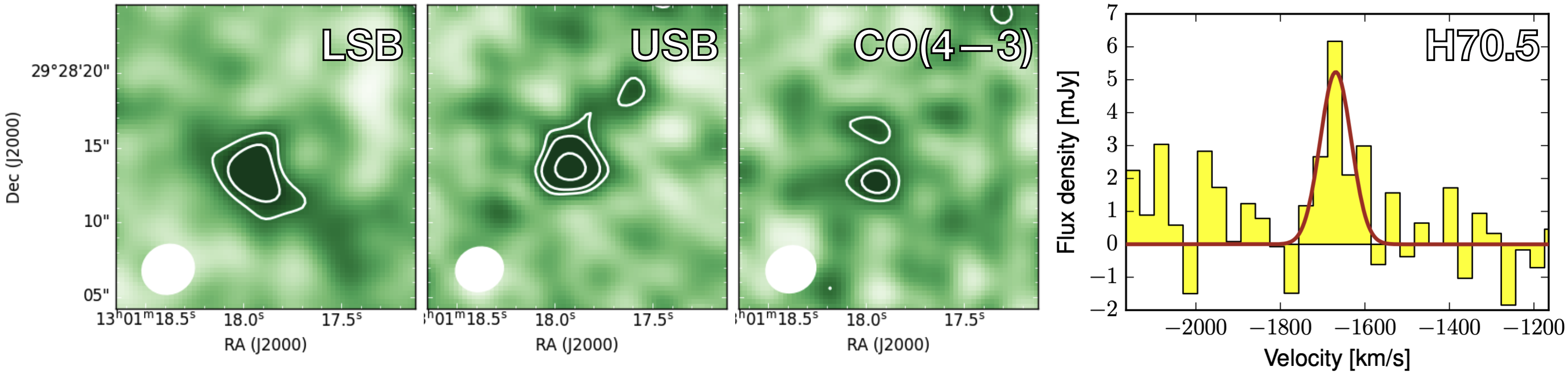}
    \includegraphics[width=0.8\linewidth]{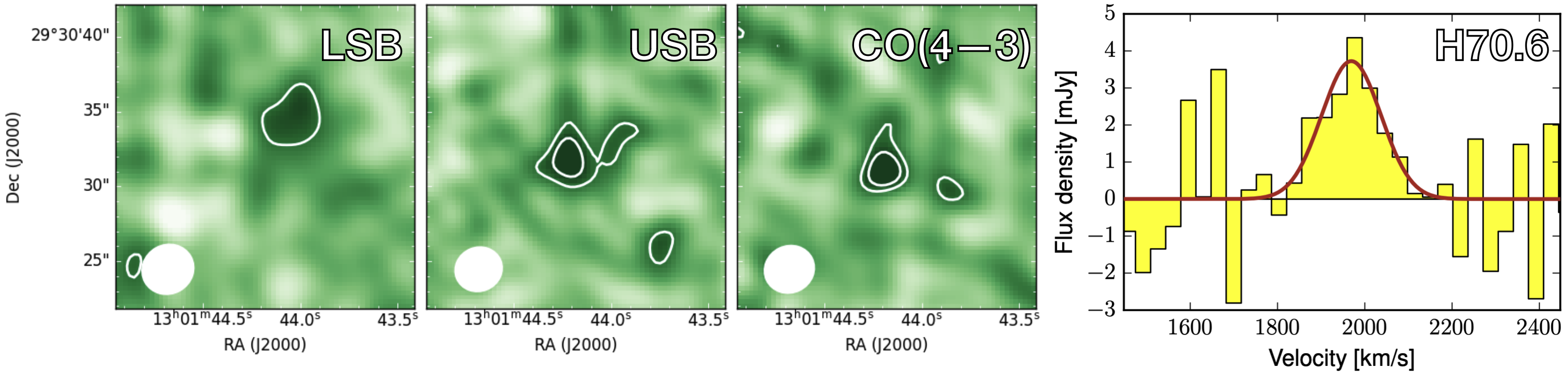}
    \caption{The continuum emission, as well as the moment-0 and spectra of the NOEMA observations of the three sources with spectroscopic line detections (H70.2, H70.5 and H70.6; from top to bottom). The poststamps are cut out at 20 by 20 arcseconds, and the contours are drawn at 2, 3, and 5 $\sigma$. The bottom left ellipse indicates the beam size. The spectra are shown in 35~km/s bins, and the velocity axis is derived relative to the zero-velocity of HerBS-70E. 
    }
    \label{fig:SpectraOfHerBS702}
\end{figure*}

Table~\ref{tab:NOEMAfluxes} shows the resulting continuum and fluxes for the nine NOEMA-observed targets. In total, four sources are detected in their dust continuum, and deep $3 \sigma$ upper limits are provided for the other five sources. 
For three of the sources with continuum detections, emission lines are identified in the spectroscopic NOEMA observations, which are shown in Figure~\ref{fig:SpectraOfHerBS702}. H70.2 has a very wide emission line ($\sim 960$~km/s), while the H70.5 and .6 have much more narrow emission lines. 
None of the sources have accompanying \ci{} line emission, suggesting the observations are not deep enough to detect the atomic carbon emission. The observed-frame wavelength of the \ci{} emission of H70.6 lies outside of the observed bandwidth.

\section{Discussion}
\label{sec:discussion}
In the next sections, we present a detailed analysis of the spectroscopic observations and examine the overdensity hypothesis in light of these observations (Sect.~\ref{sec:gaspoorprotocluster}). In Sect.~\ref{sec:continuumoverdensityestimates}, we evaluate the wider environment of HerBS-70 using the continuum observations, which was part of our motivation towards the spectroscopic observations. Sect.~\ref{sec:halomass} provides a cosmological picture of the evolution of the HerBS-70 environment towards $z=0$, and we conclude with a perspective on future protocluster studies among DSFG sources (Sect.~\ref{sec:findingProtoclustersInHerschel}).

\subsection{A gas-poor protocluster core at $z = 2.3$}
\label{sec:gaspoorprotocluster}
As will be shown in the subsequent sub-section, the continuum SCUBA-2 observations provide modest estimates of the overdensity bias of HerBS-70 ($\delta_{\rm bias} \approx 3$) through a two-dimensional perspective of the environment. The continuum observations provide only a low-significance statistic on the overdensity, exacerbated by the negative-K correction that probes infrared luminosities regardless of the source redshift \citep{blain1999} which dilutes the true protocluster members with line-of-sight galaxies. Instead, the NOEMA observations are able to confirm that three additional sources are part of the HerBS-70 system, corresponding to a total additional star-formation rate of 1650~M$_{\odot}$/yr (i.e., the sum of the star-formation rates of H70.2, .5 and .6 from Table~\ref{tab:sources}). 
{\color{referee} In total, the identified system contains five dusty galaxies, and we have excluded an additional six dusty SCUBA-2 identified galaxies from the system. Finally, 10 sources remain without spectroscopic follow-up and their protocluster membership is still not known.}
{\color{referee} Of the three additional sources contributing to the overdensity bias in Figure~\ref{fig:mapOfHerBS70}, two turned out to be real protocluster members. This suggests that the continuum bias estimate ($\delta_{\rm bias} \approx 3$) is well recovered by the NOEMA observations, a surface-density bias of $\delta_{\rm bias} = 2$.
}

The cosmic star-formation rate density at $z = 2.3$ is roughly 0.063~M$_{\odot}$/yr/Mpc$^3$ \citep{madau2014,Algera2023}. Our observations probe a 14 by 14 arcminute region within $\delta z = 0.04$ ($\pm 2000$~km/s) relative to the central HerBS-70 frequency. Here, the distance probed by a $\delta z$ is equal to the difference in co-moving distances at $z = 2.29$ and $z = 2.33$ divided by $(1+z)$. The resulting volume of 735 Mpc$^3$ (i.e., 7 by 7 by 15 Mpc) is thus expected to contain star-formation of 46~M$_{\odot}$/yr. Excluding the central sources, the resulting spectroscopic overdensity is thus 36 larger than the field environment, a ten-fold increase of the overdensity estimate compared to continuum observations. If we just interpret the central region, i.e., the protocluster core identified by the SCUBA-2 observations, the star-formation in this small 1 by 1 by 15 Mpc region is 980~M$_{\odot}$/yr (i.e., H70.2 and .6). This is roughly a thousand times higher than expected from cosmic star-formation rate density arguments.

\begin{figure}
    \centering
    \includegraphics[width=\linewidth]{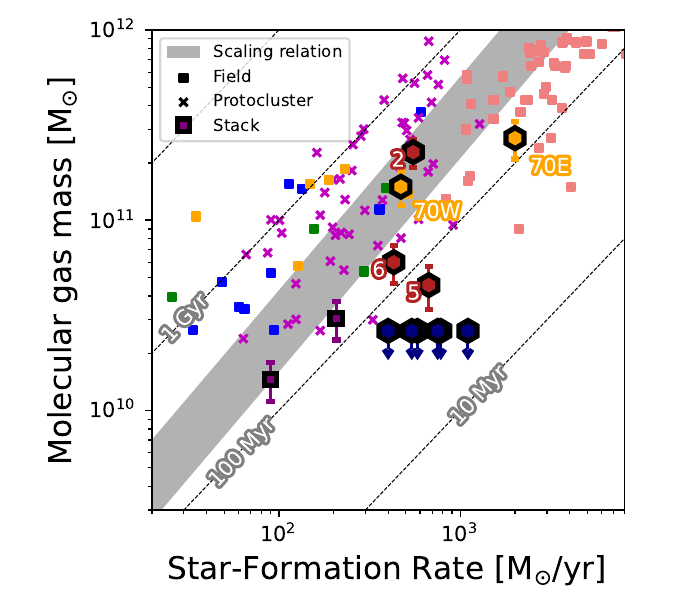}
    \caption{The molecular gas mass estimates based on the NOEMA observations for HerBS-70E, W, and the three detected sources ({\it orange and red filled hexagons}), as well as the upper limits on the six non-detected galaxies ({\it blue filled hexagons}). These are compared against scaling relations from field galaxies \citep[fill and magenta squares][]{Scoville2016}, sub-mm galaxies \citep[pink squares][]{Bendo2023,Hagimoto2023}, known protoclusters \citep[purple crosses][]{Zavala2019,PerezMartinez2023}, and stacking experiments \citep[big squares][]{Alberts2021}. The NOEMA-identified sources lie below the scaling relation, and the deep NOEMA observations further provide confidence our observations are deep enough to characterize galaxies down to depletion times $t_{\rm dep} < 80$~Myr. }
    \label{fig:MolMass}
\end{figure}

Figure~\ref{fig:MolMass} shows the molecular gas mass estimates for the NOEMA-observed sources as a function of their star-formation rate. These are compared to known field galaxies \citep{Scoville2016}, dusty star-forming galaxies \citep{Bendo2023,Hagimoto2023}, known protoclusters \citep{Zavala2019,PerezMartinez2023} and stacking experiments from Planck-identified regions \citep{Alberts2021}. The HerBS-70E and -70W molecular gas masses are based on the direct CO(1--0) observations reported in \cite{Stanley2023}. 
The molecular gas masses are derived from the CO(4--3) transition, which has intrinsic uncertainty in the conversion between CO(4--3) and the ground transition CO(1--0). We use the typical equations from \cite{solomon2005} to convert the CO(4--3) to the line luminosity, and use the CO(4--3) to CO(1--0) conversion ratio of $r_{4,1} = 0.46 \pm 0.07$ from \cite{Harrington2021}. In this paper, we assume the CO luminosity to molecular gas conversion factor of $\alpha_{\rm CO} = 4.0$~${\rm M_{\odot} (K~km~s^{-1}~pc^2)^-1}$ based on \cite{Dunne2021,Dunne2022}. For fair comparison, we correct all studies that use different $\alpha_{\rm CO}$ (typically set to 0.8~${\rm M_{\odot} (K~km~s^{-1}~pc^2)^-1}$) to the value assumed in this study. 

The scaling relation derived from the \citet{Scoville2016} sources, as well as the gas depletion timescales of distant galaxies (i.e., the available molecular gas divided by the star-formation rate) indicate that most of the sub-mm identified sources in the HerBS-70 system fall on the low end of the scaling relation with depletion times between 80 and 500~Myr. These observations seem to suggest that the protocluster system is rapidly processing its available gas, and could move towards a quenched cluster system in the $z \approx 1$ Universe \citep{Shimakawa2018}.

This low depletion timescale is also reflected in a comparison to the cosmic molecular gas density. The combined gas mass in H70.2, .5 and .6 equals a total of 33~$\pm$~6~$\times 10^{10}$~M$_{\odot}$. 
Meanwhile, the cosmic molecular gas density estimate in \citet{Decarli2019} predicts $\sim 5 \times 10^7$~M$_{\odot}$/Mpc$^{3}$. The gas density across the 735 Mpc$^3$ volume surrounding the HerBS-70 environment probed by these observations is roughly only nine times higher than the average gas density in the $z = 2.3$ Universe. Although CO(4--3) might not represent the total gas mass in the HerBS-70 system, this could point to a rapidly-quenching system at $z = 2$, in line with the evolutionary picture reported in \cite{Shimakawa2018}, where clusters move from star-forming to quenched in the $z = 2 - 3$ Universe as their gas reservoirs deplete.

\subsection{Continuum overdensity estimates}
\label{sec:continuumoverdensityestimates}
The SCUBA-2 observations revealed twenty-one $> 3.5 \sigma$ dusty sources in the vicinity of HerBS-70 (excluding the central source). Here, we analyse the continuum properties of the SCUBA-2 identified sources, and evaluate the overdensity hypothesis based on these continuum images and multi-wavelength continuum data. Although the spectroscopic observations are able to better verify the nature of these continuum sources, a thorough investigation of the continuum observations can provide context for subsequent sub-mm camera observations that have not yet had the luxury of spectroscopic follow-up.

\subsubsection{Infrared photometric redshifts}
\label{sec:herschelcounterparts}
Eleven sources identified by SCUBA-2 (excluding HerBS-70E\&W) have nearby {\it Herschel} galaxies within one JCMT beam identified through the HELP catalogue \citep{Shirley2021HELP}. This {\it Herschel} catalogue is produced using source extraction from the 250, 350 and 500~\micron{} maps, and thus includes sources that would have otherwise been too faint to be detected at 250~\micron{} alone, which is standard for the other H-ATLAS catalogues \citep[i.e.,][]{valiante2016,furlanetto2018}. 
% We include {\it Herschel} sources within one beam of the JCMT ($\sim 13$ arcseconds). 
% The number of SCUBA-2 sources with {\it Herschel} counterparts decreases with decreasing flux, which is most likely due to a mixture of false-positives (i.e., contaminants) and lower associated {\it Herschel} fluxes. 
Interestingly, in addition to HerBS-70, only three of the six sources detected at $>5 \sigma$ at 850~\micron{} have {\it Herschel} counterparts (with 5 {\it Herschel} detections at 250~\micron{} for the 11 SCUBA-2 sources detected at $>4 \sigma$). As shown in Appendix A (Fig. \ref{fig:herschelphotometry}), the reason for the lack of {\it Herschel} photometry is either due to nearby bright galaxies that affect the source extraction or because the relatively faint 250~\micron{} flux densities at which the sources are extracted \citep{valiante2016}, which can be expected for $z = 2.3$ sources. %This suggests that the 850~\micron{} data point is not far down the Rayleigh-Jeans slope, and these sources are at (similarly-)high redshift as HerBS-70, as is discussed in Subsection~\ref{sec:photoz}.

% {\bf CHANGE THIS TEXT}
For those SCUBA-2 sources without any {\it Herschel} counterparts, we extract the SPIRE fluxes from the calibrated, background-subtracted images from the second data release of H-ATLAS \citep{smith2017}\footnote{\url{https://www.h-atlas.org/public-data/download}}. We use the peak SCUBA-2 position, and extract the fluxes at 250, 350 and 500~\micron{} directly from these maps. We use the map-based average noise estimates of 6.1, 5.6 and 6.0~mJy for 250, 350 and 500~\micron{}, respectively. %The HELP catalogue does not contain {\it Herschel} sources near the positions of sources H70.1, .2, and .5, . The lack of catalogue entries at the positions of these sources is thus likely due to source confusion effects. The photometric redshift fits based on the fluxes directly extracted from the map are in line with $z_{\rm phot} = 2.3$.
We note the possibility of a boosted 500~\micron{} flux due to the bright \textsc{[C~ii]} emission line at an observed wavelength of $\sim 520$~\micron{} \citep{Smail2011,Seymour2012,Dannerbauer2014}. This effect would boost the photometric redshift estimates of the sources, although the expected deviation would depend on the inter-stellar medium properties of each galaxy, and are hard to quantify without dedicated observations \citep{Burgarella2022}.

% \subsection{Near-infrared counterparts}
% We cross-compare the positions of the 850~\micron{} sources against existing catalogues of UKIRT InfraRed Deep Sky Surveys (UKIDSS; \citealt{Lawrence2007}, using the Large Area Survey data from the {\sc UKIDSSDR11PLUS} release) and find a marginal detection at the position of HerBS-70E of K$_S = 18.49$~mag$_{AB}$ at $\sim 1$~arcsecond from the peak NOEMA position, with no counterpart in Y, H nor J down to $\sim 18.5$~mag$_{AB}$ shown in Figure~\ref{fig:photometry}. We follow the method described in \citealt{Kodama1999} and \citealt{Koyama2013} to transform this K$_S$ band flux into a rough estimate of the stellar mass, assuming a formation redshift of $z_f = 5$, giving $M_* \approx 5 \times{} 10^{11}$~M$_{\odot}$. This large stellar mass places it at the high end of the star-forming main sequence relation. Instead, there could also be significant contribution from the AGN component (as seen in radio; Stanley et al. subm), or a contribution of H$\alpha$ within the K$_S$ band. %In Subsection~\ref{sec:bcg} we provide more detailed estimates of the multi-wavelength nature of HerBS-70E.
% Only two other galaxies (36 and 39) have UKIDSS counterparts within 3~arcseconds, with suggested stellar masses of 5 and 22~$\times{} 10^{12}$~M$_{\odot}$, which seem unrealistic and are likely chance alignments. We further discuss the near-infrared detection of HerBS-70E in Subsection~\ref{sec:bcg}.

\begin{figure}
    \includegraphics[width=\columnwidth]{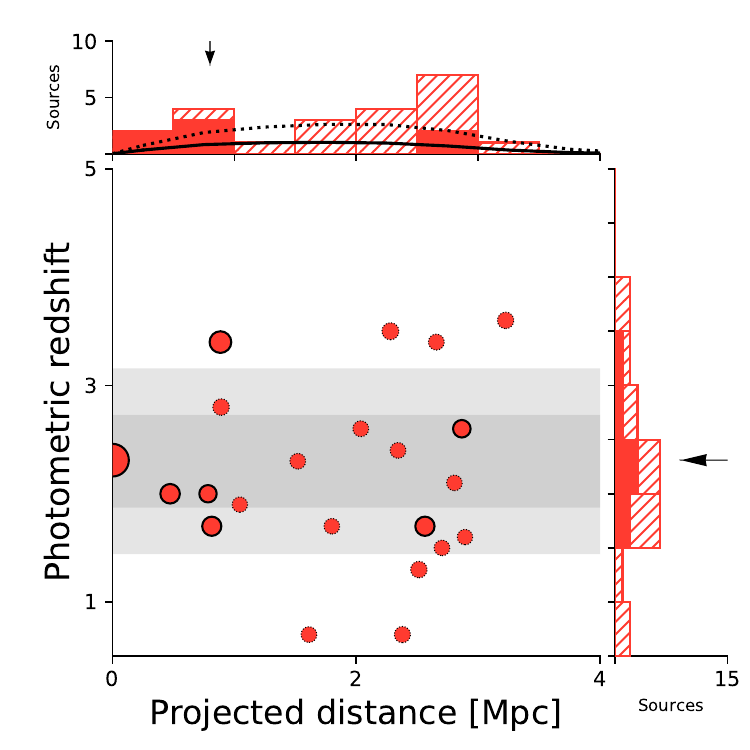}
    \caption{The projected distance from the central source (HerBS-70) versus the photometric redshift estimate from a fit of the template in \citet{pearson13} shows an excess of $> 5\sigma $ sources within one megaparsec. Larger {\it red dots} refer to higher signal-to-noise detections, where sources with a {\it black border} have at least SNR greater than 5. The distribution of the photometric redshift estimates of these sources peaks at $z_{\rm phot}=2.5$, with the filled histograms showing all $>5\sigma$ sources. The arrow on the right-hand-side histogram indicates the $z_{\rm spec} = 2.3$ of HerBS-70, and the arrow along the top histogram shows the expected size of 0.8 Mpc for a protocluster core at $z=2$  \citep{Chiang2017}. The {\it grey filled regions} indicate the 1 and $2 \sigma$ uncertainty in the photometric redshift relative to $z = 2.3$. The majority of robustly-detected ($5 \sigma$) sources lie around the photometric redshift of HerBS-70, and fall within the expected size of a protocluster core. The {\it dashed} and {\it solid} lines in the top histogram show the expected distribution of sources based on the number counts from \citet{Simpson2019}.
    A sizeable fraction of sources lie close to HerBS-70, suggesting an overdensity of 3.5 times the number of sources expected for field galaxies.}
    \label{fig:redshifts}
\end{figure}

% \subsection{Infrared photometric redshifts}
% \label{sec:photoz}
We fit the far-infrared fluxes from both the {\it Herschel} and SCUBA-2 photometry, using a two-temperature modified black-body from \cite{pearson13}. This template is suitable, since it was derived from {\it Herschel} photometry for 40 galaxies at low- and high-redshift. % For sources that do not have an identified {\it Herschel} counterpart in the HELP catalogue, we extract the fluxes directly from the calibrated background-subtracted images from \cite{smith2017}. 
Figure~\ref{fig:redshifts} shows the resulting redshift distribution for all 22 sources (incl. HerBS-70EW) identified by SCUBA-2 as a function of their projected distance to the central HerBS-70 position. The distribution of photometric redshift estimates lie around $z_{\rm phot} = 2.5$, which encompasses the spectroscopic redshift of HerBS-70 ($z = 2.3$, {\it horizontal arrow}). We note that this is not necessarily an indication of an overdensity, as the typical redshift distribution for {\it Herschel}- and SCUBA-2 extracted sources is around $\bar{z} = 2.5 \pm 0.5$ \citep{Birkin2021}. 
Previous studies have shown that the average uncertainty in a photometric redshift estimate is roughly $\Delta z = 0.13 (1+z)$ \citep{pearson13,ivison16,bakx18}. As an extra check, we also compute the photometric redshifts using the method described in \cite{ivison16}, and find broad agreement between the photometric redshift estimates. We show the fitted spectra and poststamps of the six $> 5 \sigma$ sources in Appendix Figure~\ref{fig:herschelphotometry}.

While the photometric redshift distribution of SCUBA-2 sources is unable to identify an overdensity of sources, the projected distance between each SCUBA-2 source and HerBS-70 suggests a marginal overdensity. The top histogram in Figure~\ref{fig:redshifts} shows the distance distribution of sources, where the observed number of sources (filled and hatched for 5 and $3.5 \sigma$ sources, resp.) lie substantially above the predicted number of sources (solid and dashed lines for 5 and $3.5 \sigma$ sources, respectively).
The predicted number of sources are calculated based on the number counts from \cite{Simpson2019}. For each pixel in the noise map, we calculate the expected number of sources above $3.5$ and $5 \sigma$ as a function of the radial distance. We indicate the expected size of the protocluster core ($r = 0.8$~Mpc; \citealt{Chiang2017}) with a {\it vertical arrow}.
At the shorter distances, the number of $5 \sigma$ sources is in noticeably above the expected number of sources.
At longer distances, the number of $3.5 \sigma$ sources appears to exceed the expected number of sources, although this could be inflated by the number of {\it false positives}.

\subsubsection{Source counts}
\label{sec:SourceCounts}
We calculate the number counts, $N (>S')$, by counting the number of sources above a certain deboosted flux $S'$ across the area, $A$, in which they would be detected. We account for the non-flat noise profile, and calculate the area where we would be able to detect sources at a given flux above 3$\sigma$.  We account for the false-positive ($FP$; contamination) and false-negative ($FN$; completeness) rates using 
\begin{equation}
    N(>S') = \sum_{\forall S_i > S'}  \frac{FP}{FN \times{} A}. \label{eq:sourcecounts}
\end{equation}
The noise estimates on the source counts are proportional to the square root of the number of sources contributing to the source count estimate, where the fluxes are accounted for Eddington boosting.

\begin{figure}
    \includegraphics[width=\columnwidth]{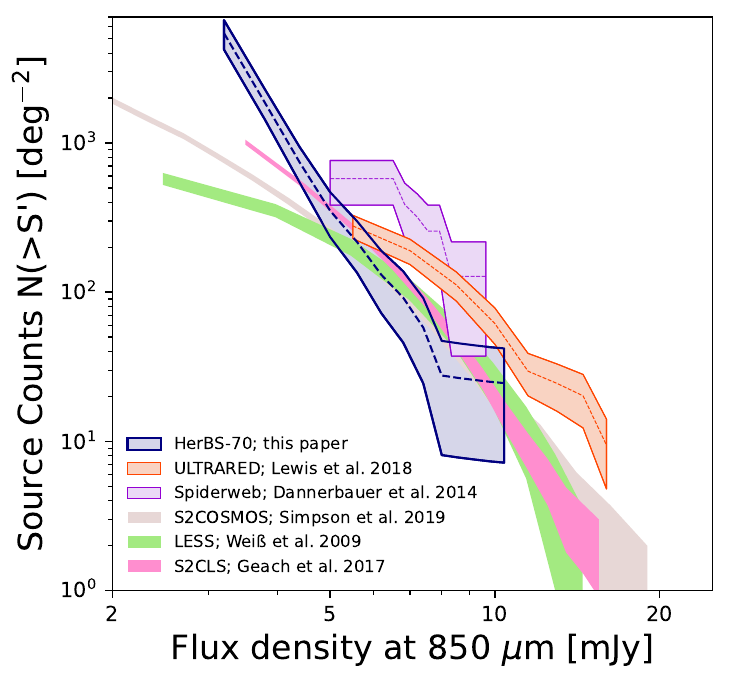}
    \caption{The source counts at 850~\micron{} show the number of sources above the normalized flux per square degree. The source counts are compared against field surveys \citep{weiss2009,Casey2013Scuba2,geach2017,Simpson2019} and known protocluster systems \citep[][shown with rimmed fill]{Dannerbauer2014,Lewis2018}. The source counts of HerBS-70 are in line with the bulk of field galaxies at flux densities from 4 to 10~mJy, and suggest moderate overdensity at the faint- and bright ends of the source counts. As previously found in \citet{Lewis2018}, source counts struggle to identify overdense regions, and we suggest instead using spatially-resolved studies to look for cosmic overdensities.
    }
    \label{fig:sourcecounts}
\end{figure}

Figure~\ref{fig:sourcecounts} shows the source counts of the HerBS-70 SCUBA-2 observations against field surveys \citep{weiss2009,Casey2013Scuba2,geach2017,Simpson2019} and known protocluster systems \citep[][]{Dannerbauer2014,Lewis2018}. %Similar to the \cite{Dannerbauer2014}, we include the central source in our source count estimates. 
The central source is masked similar to other studies (e.g., \citealt{Lewis2018}) %, and the corresponding consequence is a lower estimate for the source count, that is truncated at 10.9~mJy, i.e., the corrected flux density of the second-brightest source. 
and, thus the the deep observations provide secure estimates of the source counts from roughly 10~mJy down to $\sim 2.5$~mJy. 

The source counts for the region around HerBS-70 are comparable with field galaxies over the majority of the flux density regime. The faint end of the source counts appears to diverge from all models, due to the deep observations and the shrinking area where our observations probe below 3~mJy. The bright end of the source counts is in line with the behaviour seen for moderately-overdense regions such as the ones seen in \cite{Lewis2018} and \cite{Lacaille2019}, who also struggle to find evidence for overdensities in the source counts alone. While our analysis on particular regions in the SCUBA-2 map indicated strong overdensities (e.g., a factor of 2.9 seen in the centre of Figure~\ref{fig:mapOfHerBS70}; see Section~\ref{sec:protoclustercore}), such details appear to be averaged out across the wider map as suggested by the source counts \citep[c.f.,][]{Zhang2022}. 

% \section{Overdensity of sub-mm sources}
\subsubsection{An overdensity of continuum sources}
\label{sec:protoclustercore}
As shown in the previous sub-sections, the photometric redshifts and source counts derived from the 850 \micron{} continuum map together with additional continuum data do not indicate clear evidence for an overdensity in the wide Mpc-scale environment around HerBS-70. 
However, as shown in Figure~\ref{fig:redshifts}, there is an indication of a source overdensity in the centre of the HerBS-70 field.
We estimate the overdensity of sources close to HerBS-70 based on a direct comparison to the source counts from \cite{geach2017}, and we visualize that using blue contours in Figure~\ref{fig:mapOfHerBS70}. We generate these contours for galaxies detected above $5 \sigma$ using a method that compares the expected number of sources against the actually-observed distribution of sources, smoothed by a kernel on the size of the typical protocluster core \citep{Chiang2017}.

We calculate the expected number of sources using the following method.
We use the noise map to calculate the $5 \sigma$ flux limit for each individual pixel, and use the source count model from \cite{geach2017} to calculate the per-pixel expected number of sources (a value much smaller than 1). These are the `expected background' against which we observe the true detected sources. Since these are the `expected' number of sources, we reflect them as negative values, and for each source identified above $5\sigma$, we replace the index by a value with a 1 (since one source was actually observed at this single pixel). Note that we exclude the central source HerBS-70EW from this calculation. We then smooth the resulting image down to the scale of a typical overdensity (here we take a Gaussian kernel with a characteristic width, $\sigma$, of 1 arcminute (500~kpc), or a full-width at half-maximum of 1.2~Mpc, roughly matching the expected size of a protocluster core at $z = 2$ \citep{Chiang2013}. As seen in Figure~\ref{fig:mapOfHerBS70}, the central region appears to be overdense by a factor of $\delta_{\rm bias} = 2.9$. 

This region features three $> 5 \sigma$ sources on top of the HerBS-70 system, and as shown in previous research \citep{Lacaille2019}, these low-number statistics result in a high uncertainty on the bias. We estimate the uncertainty in the bias by repeatedly re-drawing from our $> 5 \sigma$ sample, and calculating the bias value within 1~Mpc of HerBS-70. The resulting bias is $\delta_{\rm bias} = 3.4 \pm 1.3$. We compare this estimate against known bright sources from fields without obvious sub-mm overdensities. This is important, since sub-mm galaxies are known to cluster together, even if they are not associated to a protocluster system. In this case, we use the UDS field from the COSMOS Legacy Survey \citep{geach2017} as a comparison field, similar to previous protocluster studies \citep{Lacaille2019}. 
We examine the spatial distribution of $> 5 \sigma$ sources ($N = 335$) in the 50 arcminute diameter UDS field by redrawing this selection, and calculate the observed bias in the environment around a randomly-chosen $> 5 \sigma$ source. This central source is subsequently masked -- same as with our HerBS-70EW analysis -- and the peak bias within 1~Mpc is then evaluated. The average bias is $\delta_{\rm bias} = 0.7 \pm 2.2$. As a sanity check, we find that the average bias across the entire 50 arcminute diameter UDS field is $\delta_{\rm bias} = 0.0 \pm 1.5$. The mean of zero is reasonable\footnote{Note that the uncertainty is not the uncertainty on the mean, but instead the field-to-field variation.}, since we derived the expected number of sources from the \cite{geach2017} source counts. Similarly, the fact that the typical bias is around 1 is in line with two-point correlation studies for sub-mm sources, which find a typical spatial association of $w(\omega = 1') = 1$ \citep{Amvrosiadis2019}. If we directly compare a random draw of the UDS against a resampled HerBS-70 field, we find a higher bias in HerBS-70 for 90 per cent of all draws. Similar to the relatively poor statistics of previous continuum camera studies (e.g., \citealt{Lewis2018}), we find a tentative indication for a HerBS-70 overdensity, however note that a selection effect (i.e., spatial clustering of bright sub-mm sources) cannot be ruled out on continuum data alone (e.g., \citealt{Chen2023}).

% The photometric redshifts suggest that most of the six $5 \sigma$ sources reside in the same environment as HerBS-70. 
Evaluating the bias of HerBS-70, particularly the inner 1~Mpc shows a strong overdensity relative to the field \citep{Simpson2019}, as seen in Figures~\ref{fig:mapOfHerBS70} and \ref{fig:redshifts}. This suggests that at some of the $5 \sigma$ detected galaxies could be true protocluster core members. %, of which all sources show faint 150~MHz emission in the LOFAR data (Figure~\ref{fig:WISE_overdensity}).
The overdensity bias changes with the size of the smoothing kernel, with tests at smaller and larger kernels resulting in smaller biases. This agrees with the distribution of the projected distances, which also suggests that the potential HerBS-70 system would be $\sim 1$~Mpc in size, and lines up with models and past observations. For example, \cite{Chiang2017} show that while protoclusters collapse to form dense environments with cosmic time, the cores of protoclusters grow in size. 
Studies at higher redshift (e.g., \citealt{Oteo2018, Long2020, Remus2022} at $z = 4$) that find smaller protocluster cores on the order of $\sim 300$~kpc, while environments at the peak of cosmic star-formation \citep{madau2014} extend over larger areas, similar to the $\sim 1.2$~Mpc size suggested by our experiment, and the 0.8~Mpc size predicted by \cite{Chiang2017} at $z = 2$. %Other numerical studies have shown that the typical size is also dependent on the protocluster mass \citep{Muldrew2015}, although here we take the typical size of a protocluster core found by models of \cite{Chiang2017} to be roughly 0.8~Mpc. 

\subsubsection{The large-scale environment around HerBS-70}
\label{sec:largescaleenvironmentaround}
\begin{figure}
    \centering
    \includegraphics[width=\linewidth]{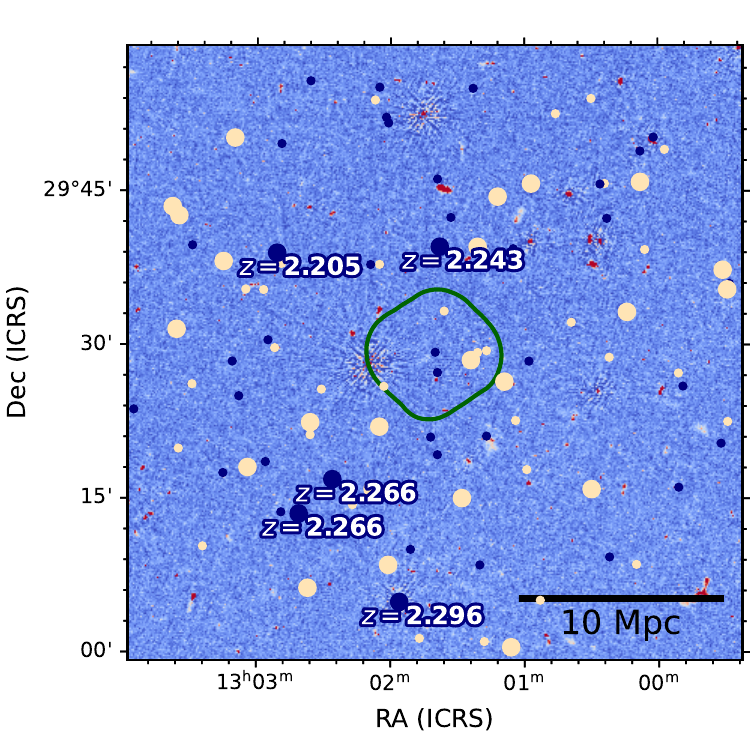}
    \caption{The spatial distribution of WISE-detected sources which are shown relative to the observed SCUBA-2 map ({\it green contour}), with the LOFAR image on the background \citep{Shimwell2022}. Sources with photometric ($\Delta z = 0.5$; {\it small marker}) and spectroscopic redshifts ($\Delta z = 0.1$; $\pm 5000$~km/s; {\it large marker}) close to the spectroscopic redshift of HerBS-70 ($z = 2.31$) are shown in {\it blue}, and sources outside of this redshift window are shown in {\it light orange}. Five quasars with spectroscopic redshifts within the region of HerBS-70 are seen out to distances of $\sim 13$~Mpc, which is around the expected size of an overdensity based on both $N$-body simulations and semi-analytical models discussed in \citet{Chiang2017}. The {\it green contours} show the extent of the SCUBA-2 observations where the per-beam source sensitivity is 2~mJy. These sources are important for follow-up observations, as they could play a role on the large-scale environment around HerBS-70, although currently not enough evidence exists to identify these sources as part of a cosmic overdensity.
    % {\it Middle panel: }A zoom in on the central region covered by SCUBA-2. The SCUBA-2 detected sources are shown as {\it black contours}, and they overlap with LOFAR-bright sources.
    % {\it Right panels: }Individual poststamps of all SCUBA-2 detected sources with significant LOFAR detections. Specifically, 150~MHz emission is seen for HerBS-70E and W, as well as sources 1, 2, 3, 4, 6, 11, 13, 14, 19, and 20. 
    }
    \label{fig:WISE_overdensity}
\end{figure}

In an effort to explore the larger-area surroundings of HerBS-70 where most of the protocluster environment should be located \citep{Chiang2017}, we compare against the catalogue of WISE sources\footnote{https://wise2.ipac.caltech.edu/docs/release/allsky/expsup/sec1\_4b.html} from the WISE All-Sky catalogue \citep{Wright2010,Cutri2012} with 30~arcmin around HerBS-70, and compare the positions against data from the recent LOFAR release \citep{vanHaarlem,Shimwell2022}. The WISE sources will likely be sensitive to ordinary, massive galaxies at lower redshifts, while the 150~MHz LOFAR data is sensitive to radio-bright galaxies and quasars.\footnote{The LOFAR maps are produced following the method described in \cite{Sweijen2022}, using the image archive at \url{https://lofar-surveys.org/dr1_release.html}.} The LOFAR sensitivity is better than 0.1~mJy/beam at $\sim 6$~arcsec resolution. 
The {\it blue points} in Figure~\ref{fig:WISE_overdensity} mark galaxies and quasars within $\Delta z = 0.25$ for photometric redshifts ({\it small points}) and 0.05 for spectroscopic redshifts ({\it large points}, equivalent to $\pm 5000$~km/s). A bright source to the east of HerBS-70 has been masked out, although some radially-extending artefacts still remain in the image that could affect sources in the eastern side of the SCUBA-2 map. Sources outside of these redshift windows are shown in {\it light orange}. Five quasars with known spectroscopic redshifts \citep{Berger1977,Richards2009,Richards2015,Paris2018} are seen out to projected distances of $\sim 13$~Mpc from HerBS-70, which is around the expected size of an overdensity based on both $N$-body simulations and semi-analytical models at $z = 2$, as discussed in \cite{Chiang2017}. 
% We report 150~MHz emission associated with the sources HerBS-70E and W, as well as sources 1, 2, 3, 4, 6, 11, 13, 14, 19, 20. We note that source 14 is likely a line-of-sight confusion due to the offset between SCUBA-2 and LOFAR positions.
Currently, not enough information is available on these quasars to identify an overdensity, however follow-up of these sources might help identify the large-scale environment around the HerBS-70 system. These bright sources at larger separation could further help identify additional evolutionary pathways of quenching of star-formation, as a widespread triggering of AGN is not uncommon for protocluster environments \citep[e.g.,][]{Casey2016}.

% Combined with the five quasars with catalogued spectroscopic redshifts \citep{Berger1977,Richards2009,Richards2015,Paris2018}, there appears evidence of excess radio emission compared to the field, and a likely widespread trigger of active nuclei in the environment around HerBS-70 \cite[e.g.,][]{Casey2016}. Only one L$*$ quasar is expected in a volume of 100\,000 comoving Mpc$^3$, a reasonable lower-limit \citep{Shen2020}. The $\delta z = 0.05$ selection ($\pm 5000$~km/s) on the five quasars suggests a much higher density of five, plus the additional four SCUBA-2 identified radio-bright sources, within $\sim 25 \times{} 25 \times{} 19 = 12\,000$ Mpc$^3$, although a spectroscopic confirmation of the nearby sub-mm sources and more thorough investigation of their nuclear emission through deep radio observations are necessary follow-up steps.

\subsection{Cosmological evolution of the HerBS-70 halo}
\label{sec:halomass}
\begin{figure}
    \centering
    \includegraphics[width=\columnwidth]{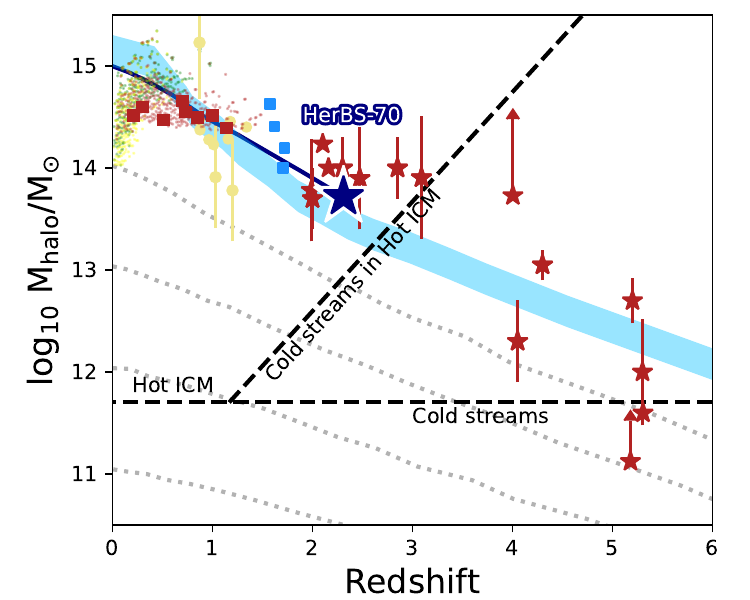}
    \caption{The halo mass distribution of HerBS-70 ({\it blue star}) is shown against various samples of clusters and models of the evolution of clusters with redshift. {\it Red stars} refer to cluster-masses derived from sub-mm camera observations \citep{Steidel1998,Kurk2000,Blain2004,Chapman2009,Daddi2009,Lehmer2009,Tamura2009,Capak2011,Kuiper2011,Walter2012,Hodge2013,Dannerbauer2014,Yuan2014,Casey2015,Casey2016,Chiang2015,Diener2015,Umehata2015,Miller2018,Lacaille2019,ConnorSmith2019,Hill2020,Long2020,Calvi2021,Guaita2022}. The {\it yellow circles} at $z = 1$ are from the GCLASS survey \citep{vanderBurg2014}, and the {\it blue squares} represent high-redshift virialized clusters \citep{Stanford2012,Zeimann2012,Mantz2014,Newman2014}.
    At the $z < 1.5$ Universe, we represent the vast samples of local clusters ({\it pink points}) compiled by \citet{Bleem2015} and references therein using {\it red squares} for clusters from SPT, {\it green pixels} for clusters from Planck, and using {\it yellow pixels} for clusters from the ACT.
    The scaling relations ({\it dashed lines}) are from the Millennium I and II simulations \citep{McBride2009,Fakhouri2010} shown at present-day masses of $10^{11}$ to $10^{14}$~M$_{\odot}$. We scale the evolution of the most massive halo to the $z = 2.3$ mass of HerBS-70 to forward-model the cluster redshift into the present age ({\it blue line}), finding a total mass on the order of $\sim 1 \times{} 10^{15}$~M$_{\odot}$. The {\it filled blue} region represents the evolution of the most massive cluster (+ errors) from \citet{Chiang2013}. 
    The {\it thick dashed black lines} indicate the redshift-halo mass regions where the mechanics of gas inflow and the cooling mechanisms of galaxy clusters transition \citep{Dekel2006,Overzier2016}. This causes the subsequent quenching of galaxy clusters into local sources \citep{Shimakawa2018}.
    }
    \label{fig:halomass}
\end{figure}

Figure~\ref{fig:halomass} shows the evolutionary pathway of the HerBS-70 system compared to known clusters and protoclusters and evolutionary models. High redshift cluster mass estimates offer a cursory glimpse at the potential evolutionary track of systems in the early Universe, although they are dependent on multiple assumptions and scaling relations, which results in near-order of magnitude uncertainties in the halo mass estimates. In this case, we use the conservative limits of the scaling relations, in order to not over-estimate the cluster mass, although future observations should focus on improving the robustness of these estimates through spectroscopic confirmation of the members of the HerBS-70 system.

We derive the halo mass of HerBS-70 from the dust mass estimates for sources in the protocluster core (i.e., HerBS-70E \& W, and sources H70.2 and H70.6), since the SCUBA-2 observations are sensitive to the most dusty, star-forming galaxies. We focus on just the central core, since our SCUBA-2 observations have provided the best constraints on the protocluster core, and the statistics of the overdensity are averaged out towards larger scales. The combined dust masses of all sources identified in Table \ref{tab:sources} add to $8.3 \times{} 10^9$~M$_{\odot}$. We then assume a gas-to-dust ratio of 100 to acquire a total gas mass estimate \citep{Hagimoto2023}.\footnote{This approach is in line with the sum of the gas mass found in Section~\ref{sec:gaspoorprotocluster}, and should better account for the total gas mass instead of just the molecular gas mass.} Particularly, this value has a moderate evolution with redshift (see \citealt{Peroux2020} for a discussion) 
and represents solar metallicity systems \citep{Pantoni2019,deLooze2020, Graziani2020, Millard2020}, which could be a relatively good tracer of even the high-redshift Universe \citep{Vijayan2019, Litke2022,Popping2022}. 
Cluster evolution models \citep{Chiang2013} suggest that 30 per cent of the mass of the cluster is contained within the inner 0.8~Mpc, and we assume a fixed baryonic-to-dark matter fraction of 5 per cent \citep{Behroozi2018}. Based on these assumptions, the estimated total cluster mass is $6 \times{} 10^{13}$~M$_{\odot}$. %In contrast to other submm-observed protoclusters at high-redshift \citep[e.g.][]{Long2020}, this source is not in contradiction with $\Lambda$-CDM models \citep{Harrison2013}.

The evolution models from \cite{McBride2009} and \cite{Fakhouri2010} predict a future halo mass in excess of 10$^{15}$~M$_{\odot}$. Although the models are derived for different halo masses, the evolution between them is only minor, and we thus feel comfortable to scale the most massive model to the estimated halo mass of $5 \times{} 10^{13}$~M$_{\odot}$ at $z = 2.3$. The halo mass is on par with massive local passive galaxy clusters such as the $\sim$~10$^{15}$~M$_{\odot}$ Virgo cluster \citep{Fourque2001}, and the HerBS-70 system might thus represent the central quenching phase in one of the most massive galaxy clusters in the Universe. {\color{referee} Here we do well to stress that direct extrapolations of the cluster masses from dust and gas-mass estimates is prone to large errors in the final mass estimate, as well as depends on ad-hoc assumptions on the content of dusty galaxies in protoclusters. Future work could better estimate the halo mass using complementary methods, such as the abundance matching technique discussed in \cite{Long2020,Calvi2021} that requires deep optical imaging, as well as better comparison to models of cosmological overdensities \citep[e.g.,][]{Chiang2017}.
}

This short expected lifetime ($\leq 70$~Myr) is consistent with the hierarchical picture of massive galaxy formation \citep{Hopkins2008}, which might be further exacerbated by the internal AGN (e.g., \citealt{Ito2022,Mountrichas2022}). 
The sudden lack of gas feeding from the environment (e.g., \citealt{Shimakawa2018} for an evolutionary picture) suggests the future evolution of the central HerBS-70 source is likely to lead to a massive red-and-dead giant elliptical galaxy by $z = 0$ \citep{Toft2014,Ikarashi2015,Simpson2017,Stach2017}. The findings reported in this paper add to the growing evidence towards this hierarchical pathway of DSFGs to quenched elliptical galaxies given the high stellar masses \citep{Hainline2011,Aravena2016}, high specific star-formation rates \citep{Straatman2014,Spilker2016,Glazebrook2017,Schreiber2018, Merlin2019} and location in an overdense region \citep{Blain2004,weiss2009,Hickox2012}.

\subsection{Finding protoclusters among multiple-DSFG systems}
\label{sec:findingProtoclustersInHerschel}
% Recent surveys with mm/submm interferometers have resolved DSFGs at high resolution. 
Large area surveys such as the ones done with the Atacama Cosmology Telescope, {\it Herschel}, {\it Planck}, and the South Pole Telescope are able to select the apparently-brightest DSFGs in the sky. Because the bright end of the luminosity function is steep, instead of finding the intrinsically-brightest galaxies, these surveys have found large numbers of gravitationally-lensed \citep{Treu2010}, less luminous galaxies \citep{negrello2010,negrello2017,bussmann13,Bussmann2015,Vieira2013,Spilker2016}. 

As surveys move towards lower-brightness sources, the fraction of unlensed sources increases \citep[][]{Cai2013,Bussmann2015,Bakx2020VIKING,Bakx2024}. 
Follow-up observations with mm/submm interferometers (NOEMA and ALMA) have resolved these DSFGs revealing, in a few cases, multiple sources at the same redshift \citep{Hodge2013,Karim2013,Zavala2015,Scudder2016,Oteo2018,neri2019}.  However, such systems remain rare with, for instance, two binary systems found in 13 sources studied in \cite{neri2019} -- including HerBS-70 -- and 8 out of 62 in the spectroscopic survey by \cite{Urquhart2022}, i.e., 15 and 13 per cent, respectively (see also \citealt{GG2019}). Multiple systems at high-redshift are prime targets to identify dusty protoclusters and recent surveys such as $z$-GAL \citep{Cox2023} have doubled the numbers of high-$z$ sources displaying genuine multiplicity, beyond what is expected from field galaxies (Bakx et al. in prep.). Here we note that identifying and studying groups of sources at $z>3.5$ is of crucial importance to test the Lambda-CDM paradigm \citep{Harrison2013}.

% These unlensed sources are often multiples, either at the same redshift or line-of-sight confusion sources . Recent spectroscopic surveys \citep[e.g.,][]{neri2019,Urquhart2022} with interferometers are now enabling us to select galaxies with multiple components at the same redshift. Preliminarily, these systems appear relatively rare in publicly-available samples, with 2 out of 13 (15 per cent; \citealt{neri2019}), and 8 out of 62 (13 per cent; \citealt{Urquhart2022}) galaxies having multiple sources at the same redshift. Future samples, such as $z$-GAL (Cox et al. in prep.), promise to double samples with such multiplicity, and could prove a good avenue for future dusty protocluster surveys with mm/submm cameras. Here we note that particularly the highest-redshift candidates (i.e., $z > 3.5$) are able to test the $\Lambda$-CDM paradigm \citep{Harrison2013}.

As shown in this study, the case of HerBS-70 at $z=2.3$ presents a clear case of multiplicity with an overdensity of submm sources. This indicates that this binary system, identified with NOEMA, lies at the core of a massive protocluster. %Spectroscopic follow-up observations of the sources found in the field surrounding HerBS-70 are key to further strengthen the identification and constrain the properties of this protocluster.  
% Spectroscopic follow-up of the candidate protocluster members in HerBS-70 is an important next step. 
The likely quenching ongoing in the central source HerBS-70E could further be studied by, for example, absorption line studies, that are able to probe the gas cycle \citep{Spilker2018,Spilker2020,Spilker2020b,berta2021,Butler2021}, and differentiate between the effect of the AGN and star-formation feedback. Finally, at $z = 2.3$, this source is an excellent target for narrow-band observations that aim for H$\alpha$ emission using near-infrared facilities (e.g., Subaru, VLT, Keck and Gemini; \citealt{Kurk2000,Koyama2013,Koyama2021}). These could also measure independently the redshifts of the bright sub-mm sources, and provide several tens of fainter protocluster components that are missed in current submm imaging surveys. %Finally, the advent of large samples of DSFGs with spectroscopic redshifts \citep{Vieira2013,weiss2013,fudamoto2017,neri2019,Birkin2021,CCChen2022,Urquhart2022} make for an enticing sample to look for overdense regions around DSFGs with this efficient narrow-band technique using existing filters. 

\section{Conclusions}
\label{sec:conclusions}
We have presented i. the 850~\micron{} SCUBA-2 imaging of the field around HerBS-70, a binary system of bright dusty star-forming galaxies at $z = 2.3$, and ii. subsequent NOEMA observations of the nine brightest SCUBA-2 targets surrounding HerBS-70 aimed to test the protocluster hypothesis. { \color{referee} Three of these nine targeted SCUBA-2 sources were found to be at the same redshift as HerBS-70, and the remaining six are excluded from the protocluster environment.} These observations, and existing multi-wavelength data, allow us to conclude the following:
\begin{itemize}
\renewcommand\labelitemi{\tiny \textbf{$\blacksquare{}$}}
    \item HerBS-70 is located at the centre of a candidate protocluster core ($\sim 0.8$~Mpc) with a volume density  $\sim 36 $ times higher than field galaxies, based on NOEMA spectroscopic observations. These observations targeted the brightest nine continuum galaxies in the HerBS-70 environment with a surface density estimates $\sim 3$ higher than field galaxies based on the SCUBA-2 sub-mm continuum map.  %The central region to a thousand if we focus on just the central protocluster core region.
    \item Meanwhile, the continuum overdensity estimates based on number counts and across the wider continuum map do not provide evidence towards the overdensity hypothesis, indicating the limits of continuum observations in identifying overdensities. 
    \item The depletion timescales of HerBS-70E, 70W, H70.2, H70.5 and H70.6 are around 500 to 100~Myr, indicating there is rapid gas depletion, which will likely result in the rapid quenching of the environment.  %.averages out in the number counts and across the wider continuum map.%, however the small (SCUBA-2; DSFGs) and large-area (LOFAR; quasars) imaging suggest an overdense system. There is thus a likely need of a cluster-wide trigger to explain the excess of DSFGs and quasars.
    \item Forward-modeling of the dust-based halo mass estimate ($\approx 5 \times{} 10^{13}$~M$_{\odot}$) suggests a present-day halo mass of $1 \times{} 10^{15}$~M$_{\odot}$, in excess of the local, massive Virgo cluster, {\color{referee} although we note that this method is prone to biases and large errors}. 
    %Multi-wavelength analysis of HerBS-70E suggests will soon be quenched (gas fraction $< 30$~per cent). Multiple quenching pathways (AGN, gas depletion, neighbouring galaxies) exist, and dedicated follow-up observations are needed to disentangle their relative contributions towards quenching.
\end{itemize}
% Future spectroscopic imaging is needed to definitively confirm the protocluster nature of the HerBS-70 system, and any source confirmations will vastly boost the overdensity bias and halo mass estimates, either using optical or sub-mm observations.

The SCUBA-2 map and subsequent NOEMA follow-up have revealed and verified an overdensity of dusty sources around HerBS-70. The $z = 2.3$ redshift of this source makes it an excellent candidate for H$\alpha$ narrow-band follow-up \citep[e.g.,][]{Koyama2021}, in order to reveal the more numerous, but less dusty components of the HerBS-70 system. Increasing the number of protocluster components can then provide insights into the structure and properties of more normal galaxies within the HerBS-70 system.
%Many questions still remain, however, requiring follow-up observations for definitive answers. 

The advent of large spectroscopic samples of DSFGs (including sizeable numbers of multiple sources) further entice similar submm camera studies, particularly in light of the production of improved mm/submm cameras, such as NIKA2 \citep{NIKA2}, TolTEC \citep{TolTec2020}, and SCUBA-3\footnote{https://www.eaobservatory.org//jcmt/wp-content/uploads/sites/2/2019/10/Guide-to-the-new-850um-MKID-camera-performance.pdf}. Such observational studies will enable to better constrain and understand the properties and evolution of protoclusters and in the early Universe, in particular at the peak of cosmic evolution.

\section*{Acknowledgements}
{\color{referee} The authors kindly thank the anonymous referee for their comprehensive comments that significantly improved the scientific scope of this manuscript. }
TB, MH and YT acknowledge funding from NAOJ ALMA Scientific Research Grant Numbers 2018-09B and JSPS KAKENHI No.~17H06130, 22H04939, and 22J21948.
The James Clerk Maxwell Telescope is operated by the East Asian Observatory on behalf of The National Astronomical Observatory of Japan; Academia Sinica Institute of Astronomy and Astrophysics; the Korea Astronomy and Space Science Institute; the National Astronomical Research Institute of Thailand; Center for Astronomical Mega-Science (as well as the National Key R\&D Program of China with No. 2017YFA0402700). Additional funding support is provided by the Science and Technology Facilities Council of the United Kingdom and participating universities and organizations in the United Kingdom and Canada. Additional funds for the construction of SCUBA-2 were provided by the Canada Foundation for Innovation.
SS was partly supported by the ESCAPE project; ESCAPE -- The European Science Cluster of Astronomy and Particle Physics ESFRI Research Infrastructures has received funding from the European Union’s Horizon 2020 research and innovation programme under Grant Agreement No.~824064. SS also thanks the Science and Technology Facilities Council for financial support under grant ST/P000584/1. SU would like to thank the Open University School of Physical Sciences for supporting this work.
HD acknowledges financial support from the Agencia Estatal de Investigación del Ministerio de Ciencia e Innovación (AEI-MCINN) under grant (La evolución de los cíumulos de galaxias desde el amanecer hasta el mediodía cósmico) with reference (PID2019-105776GB-I00/DOI:10.13039/501100011033) and acknowledge support from the ACIISI, Consejería de Economía, Conocimiento y Empleo del Gobierno de Canarias and the European Regional Development Fund (ERDF) under grant with reference PROID2020010107. This work benefited from the support of the project Z-GAL ANR-AAPG2019 of the French National Research Agency (ANR).
LOFAR data products were provided by the LOFAR Surveys Key Science project (LSKSP; https://lofar-surveys.org/) and were derived from observations with the International LOFAR Telescope (ILT). LOFAR \citep{vanHaarlem} is the Low Frequency Array designed and constructed by ASTRON. It has observing, data processing, and data storage facilities in several countries, which are owned by various parties (each with their own funding sources), and which are collectively operated by the ILT foundation under a joint scientific policy. The efforts of the LSKSP have benefited from funding from the European Research Council, NOVA, NWO, CNRS-INSU, the SURF Co-operative, the UK Science and Technology Funding Council and the Jülich Supercomputing Centre. A.N. acknowledges support from the Narodowe Centrum Nauki (NCN), Poland, through the SONATA BIS grant UMO-
2020/38/E/ST9/00077.

%%%%%%%%%%%%%%%%%%%%%%%%%%%%%%%%%%%%%%%%%%%%%%%%%%
\section*{Data Availability}
The data underlying this article will be shared on reasonable request to the corresponding author.



\bibliographystyle{mnras}
\bibliography{example} % if your bibtex file is called example.bib

%\begin{thebibliography}{99}
%\end{thebibliography}

%%%%%%%%%%%%%%%%%%%%%%%%%%%%%%%%%%%%%%%%%%%%%%%%%%

%%%%%%%%%%%%%%%%% APPENDICES %%%%%%%%%%%%%%%%%%%%%
{ \small
$^{1}$Department of Earth and Space Sciences, Chalmers University of Technology, Onsala Observatory, 439 94 Onsala, Sweden\\
$^{2}$Department of Physics, Graduate School of Science, Nagoya University, Nagoya, Aichi 464-8602, Japan\\
$^{3}$National Astronomical Observatory of Japan, 2-21-1, Osawa, Mitaka, Tokyo 181-8588, Japan\\
$^{4}$Institut de Radioastronomie Millimétrique (IRAM), 300 Rue de la Piscine, 38400 Saint-Martin-d’H\`{e}res, France\\
$^{5}$Instituto Astrof\'{i}sica de Canarias (IAC), E-38205 La Laguna, Tenerife, Spain\\
$^{6}$Dpto. Astrof\'{i}sica, Universidad de la , E-38206 La Laguna, Tenerife, Spain\\
$^{7}$Institut d’Astrophysique de Paris, Sorbonne Universit\'{e},UPMC Universit\'{e} Paris 6 and CNRS, UMR 7095, 98 bis boulevard Arago, F-75014 Paris, France\\
$^{8}$Instituto Nacional de Astrof\'{ı}sica, \'{O}ptica y Electr\'{o}nica, Tonantzintla, 72000 Puebla, M\'{e}xico\\
$^{9}$I. Physikalisches Institut, Universit\"at zu K\"oln, Z\"ulpicher Strasse 77, D-50937 K\"oln, Germany\\
$^{10}$Leiden Observatory, Leiden University, PO Box 9513, NL-2300 RA Leiden, Netherlands\\
$^{11}$Department of Physics and Astronomy, Rutgers, the State University of New Jersey, 136 Frelinghuysen Road, Piscataway, NJ 08854-8019, USA\\
$^{12}$Department of Physics and Astronomy, University of the Western Cape, Robert Sobukwe Road, Bellville 7535, South Africa\\
$^{13}$Institut d’Astrophysique Spatiale (IAS), CNRS et Universit\'{e} Paris Sud, Orsay, France\\
$^{14}$UK ALMA Regional Centre Node, Jodrell Bank Centre for Astrophysics, Department of Physics and Astronomy, \\University of Manchester, Oxford Road, Manchester M13 9PL, UK\\
$^{15}$Dipartimento di Fisica e Astronomia “G. Galilei”, Universit’di Padova, Vicolo dell’Osservatorio 3, I-35122, Padova, Italy\\
$^{16}$Aix-Mar seille Univer sit\`{e}, CNRS and CNES, Laboratoire d’Astrophysique de Marseille, 38, rue Fr\`{e}d\`{e}ric Joliot-Curie, F-13388 Marseille, France\\
$^{17}$Department of Physics \& Astronomy, University of California, Irvine, 4129 Reines Hall, Irvine, CA 92697, USA\\
$^{18}$School of Physics and Astronomy, Cardiff University, The Parade, Cardiff, CF24 3AA, UK\\
$^{19}$School of Physics \& Astronomy, University of Nottingham, University Park, Nottingham NG7 2RD, UK\\
$^{20}$Institute of Astronomy, University of Cambridge, Madingley Road, Cambridge CB30HA, UK\\
$^{21}$Department of Astronomy, University of Maryland, College Park, MD 20742, USA\\
$^{22}$European Southern Observatory, Karl Schwarzschild Strasse 2, D-85748 Garching, Germany\\
$^{23}$Jodrell Bank Centre for Astrophysics, School of Natural Sciences, The University of Manchester, Manchester M13 9PL, UK\\
$^{24}$Centre de Recherche Astrophysique de Lyon, ENS de Lyon, Universit\'{e} Lyon 1, CNRS, UMR5574, 69230 Saint-Genis-Laval, France\\
$^{25}$Department of Astronomy, University of Cape Town, Private Bag X3, Rondebosch 7701, Cape Town, South Africa\\
$^{26}$INAF, Instituto di Radioastronomia-Italian ARC, Via Piero Gobetti 101, I-40129 Bologna, Italy\\
$^{27}$European Southern Observatory, Alonso de Córdova 3107, Vitacura, Casilla 19001, Santiago de Chile, Chile\\
$^{28}$Joint ALMA Observatory, Alonso de Córdova 3107, Vitacura 763-0355, Santiago de Chile, Chile\\
$^{29}$Ikerbasque Foundation, University of the Basque Country, DIPC Donostia, Spain\\
$^{30}$National Centre for Nuclear Research, ul. Pasteura 7, 02-093 Warszawa, Poland \\
$^{31}$ INAF - Osservatorio astronomico d’Abruzzo, Via Maggini SNC, 64100, Teramo, Italy
$^{32}$The Observatories of the Carnegie Institution for Science, 813 Santa Barbara St., Pasadena, CA 91101, USA \\
$^{33}$Department of Physics and Astronomy, University of California, Riverside, 900 University Ave, Riverside, CA 92521, USA \\
$^{34}$School of Physical Sciences, The Open University, Milton Keynes, MK7 6AA, UK\\
$^{35}$National Radio Astronomy Observatory, 520 Edgemont Road, Charlottesville, VA 22903, USA\\
$^{36}$Max-Planck-Institut für Radioastronomie, Auf dem Hügel 69, 53121 Bonn, Germany
}

\appendix

\section{Spectral fitting using Herschel and SCUBA-2 fluxes}
\begin{figure*}
    \centering
    \includegraphics[width=0.9\linewidth]{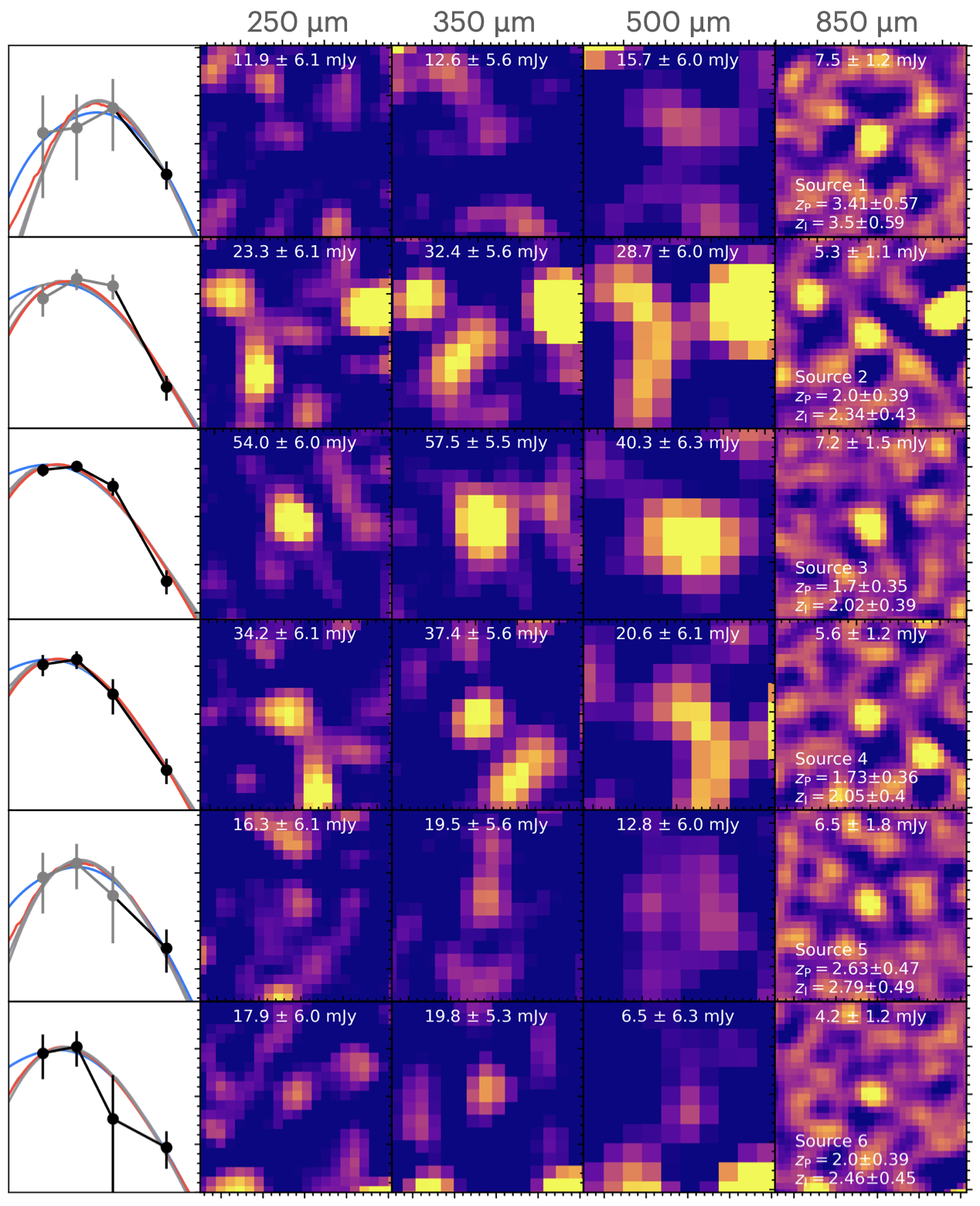}
    \caption{The left columns show the fitted spectra ({\it black markers} for detected sources and {\it grey markers} for sources without a listing in the HELP catalogue), with {\it red line} indicating the two-temperature modified black-body from \citet{pearson13} (indicated by $z_P$), {\it blue line} indicating the best-fit spectrum from the \citet{ivison16} method (indicated by $z_I$), and the grey lines showing the fitted templates with higher signal-to-noise ratios. From left to right, the images show the 250, 350, 500 and 850\micron{} maps at 1 by 1 arcminute. The photometric redshift estimates of all six sources agree with the spectroscopic redshift of HerBS-70E\&W at $z_{\rm spec} = 2.3$.}
    \label{fig:herschelphotometry}
\end{figure*}

Appendix Figure~\ref{fig:herschelphotometry} shows the fits of the six sources detected at $> 5 \sigma$ (with {\it solid black borders} in Figure~\ref{fig:redshifts}). While multiple sources are not detected in the {\it Herschel} catalogues, these observations provide strict upper limits (particularly at the wavelength of 250~\micron{}) that restrict the potential photometric redshifts of these sources. The {\it red line} shows the fitted spectrum according to the two-temperature modified black-body of \cite{pearson13}, while the {\it blue} and {\it grey lines} indicate the best-fit spectrum according to the method presented in \cite{ivison16}. Both methods appear to fit the data well.

% \section{UKIDSS data}
% \begin{figure*}
%     \centering
%     \includegraphics[width=\linewidth]{magphys.png}
%     \caption{{\it Top panel:} The Y, J, H, and K-band images from UKIDSS \citep{Lawrence2007} with the NOEMA image of HerBS-70E overlaid ({\it white contours)}. The galaxy is only seen in K-band, slightly offset from the NOEMA position. The offset is significant, beyond the expected offset from astrometric errors, and might suggest a dust-obscured and dust-emitting component within HerBS-70E.
%     {\it Middle panel:} The MAGPHYS \citep{cunha2008} model of HerBS-70E shows a well-fitted spectrum from optical to far-infrared with a fixed $z_{\rm spec} = 2.31$. The {\it blue} component shows the dust-unobscured view of the galaxy.  The residual is within $1 \sigma$. 
%     {\it Bottom panel:} The fitted parameters from MAGPHYS suggest a massive (M$_{*} \approx 10^{12.4}$~M$_{\odot}$), with a high specific star-formation rate (sSFR~$\approx 10^{-9}$~M$_{\odot}$/yr). The relatively low depletion time suggests this galaxy will rapidly quench, and likely transform into a red-and-dead massive elliptical.
%     }
%     \label{fig:photometry}
% \end{figure*}

%%%%%%%%%%%%%%%%%%%%%%%%%%%%%%%%%%%%%%%%%%%%%%%%%%

% Don't change these lines
\bsp	% typesetting comment
\label{lastpage}
\end{document}